\documentclass[epj]{svjour}
\usepackage{graphics}
\begin{document}
\title{A Discrimination Procedure between Muon and Electron in Super-Kamiokande Experiment Based on the Angular Distribution Function Method}
\author{V.I.~Galkin\inst{1} \and A.M.~Anokhina\inst{1}
\and E.~Konishi\inst{2}\and A.~Misaki\inst{3}
}                     
\offprints{}          
\institute{Department of Physics, Moscow State  University,Moscow, 119992, Russia
 \and Graduated School of Science and Technology,
Hirosaki University, 036-8561, Hirosaki, Japan
 \and Advanced Research Institute for Science and Engineering, Waseda University, 169-0092, Tokyo, Japan\\
e-mail:misaki@kurenai.waseda.jp 
}
\date{Received: date / Revised version: date}
%
\abstract{
  In the previous paper, we construct the angular distribution functions 
for muon and electron as well as their relative fluctuation 
functions to find  suitable discrimination procedure between 
muon and electron in Superkamiokande experiment. 
In the present paper, we are able to discriminate muons from
electrons in Fully Contained Events with a probability of error of
less than several \%.  At the same time, our geometrical reconstruction
procedure, considering only the ring-like structure of the Cherenkov
image, gives an unsatisfactory resolution for 1~GeV $e$ and $\mu$,
with a mean vertex position error, $\delta r$, of 5--10\,m and a mean
directional error, $\delta \theta$, of about 6$^\circ$--20$^\circ$.
In contrast, a geometrical reconstruction procedure utilizing the full
image and using a detailed approximation of the event angular
distribution works much better: for a 1~GeV $e$, $\delta r \sim$2\,m
and $\delta \theta \sim$3$^\circ$; for a 1~GeV $\mu$, $\delta r
\sim$3\,m and $\delta \theta \sim$5$^\circ$.  At 5~GeV, the
corresponding values are $\sim$1.4\,m and $\sim$2$^\circ$ for $e$ and
$\sim$2.9\,m and $\sim$4.3$^\circ$ for $\mu$.  The numerical values
depend on a single PMT contribution threshold. The values quoted above
are the minima with respect to this threshold.
Even the methodologically correct approach we have adopted, based on
detailed simulations using closer approximations than those adopted in
the SK analysis, cannot reproduce the accuracies for particle
discrimination, momentum resolution, interaction vertex location, and
angular resolution obtained by the SK simulations, suggesting the
assumptions in these may be inadequate.  
\PACS{
      {13.15.+g}{Neutrino interactions}   \and
      {14.60.-z}{leptons}
     } 
} 
\titlerunning{A Discrimination procedure in SK}
\maketitle
\section{Introduction}
\label{intro}
 In the preceeding paper \cite{r1}, we have proposed a new 
 discrimination procedure 
between electron and muon in the SK experiment instead of the standard 
SK procedure. For the purpose, we have constructed the mean angular
distribution functions for the charged particles concerned 
(muons and electrons/positrons), as well as the corresponding
 functions for the relative fluctuation.
 In the present paper, we apply our discrimination procedure to the 
SK experiment.

\section{Procedure for discriminating electron neutrinos 
from muon neutrinos: Event type definition}
\label{sec:2}
Figure~1 shows the process of classification, i.e. how event 
is called mu-like (muon event) or e-like (electron event).
 First, one needs to calculate patterns for classes
 $\mu$ and $e$ with  the help of models of Cherenkov 
light spatial-angular distribution. Second, one should compare the 
event under consideration with both patterns by calculating
 $q_{\mu}$ (7-a) and $q_{el}$ (7.b) which will be shown later
(See page~2).
Third, one compares their difference with $q_{crit}$ thus 
making a decision on the type of event.
 
In order to discriminate electron neutrino events from muon
neutrino events, we construct discrimination functions 
in the following way.

We consider each event to be a random vector 
$\mathbf{Q} \, = \{Q_j\}$ 
containing the contributions to all photomultipliers 
(PMTs) of the detector.  
Here $j$ is the PMT index: $j=1,2,...,N$ and $N$ is
the total number of PMT of the detector. There are two
classes of events to be considered by the type identification
procedure: $\omega_1 \, = \, e$ (electron initiated) and $\omega_2
\, = \, \mu$ (muon initiated). Monte Carlo simulations of event optical
images provide a possibility to study the behaviour of images of both
classes, namely, the image distribution functions
$F(Q_1,Q_2,...,Q_N;\omega_i,E_0,\mathbf{r_0},\mathbf{\theta_0})$
which are simultaneous distribution functions of PMT contributions
$Q_j$ for given particle type $\omega_i$, energy $E_0$, injection
point $\mathbf{r_0}$ and direction
$\mathbf{\theta_0}$. i.e. the most differential and 
complete characteristics of the classes.

It is, however, too expensive to deal with these functions:
in principle, one should simulate an enormous sample to know
the most differential distribution function even for
only one set of primary parameters 
$\{\omega_i,E_0,\mathbf{r_0}, \mathbf{\theta_0} \}$.
A more practical solution is to determine only the mean vector
 $\{\bar{Q_j}(\omega_i,E_0,\mathbf{r_0},
\mathbf{\theta_0})\}$
and the covariance matrix\\
$\Sigma_Q(\omega_i,E_0,\mathbf{r_0},\mathbf{\theta_0})
\, = \, cov(Q_j,Q_m)$ 
and then assume the distribution is a multi-variate normal one: \\
\\
$
p(\mathbf{Q}; \omega_i,E_0,\mathbf{r_0},
\mathbf{\theta_0})
= \,(2\pi)^{-N/2}\cdot\left(det\, \Sigma_{\omega_i}^{-1} \right )^{-1/2}\times
$
$$
\;\;\;\;\; \times exp\left\{-\left(\mathbf{Q} - \mathbf{Q^{\omega_i}} \right )^T
\Sigma_{\omega_i}^{-1} \left(\mathbf{Q}-\mathbf{Q^{\omega_i}}
\right ) \right \}, \;  \; \;\;\;\; \;\;\; (1)  
$$

Here\\ 
$
\mathbf{Q^{\omega_i}} \, = \, \mathbf{Q}
(\omega_i,E_0,\mathbf{r_0},\mathbf{\theta_0}),
\Sigma_{\omega_i} \, = \, \Sigma_Q(\omega_i,E_0,\mathbf{r_0},
\mathbf{\theta_0}).
$
\\
In this case, one still needs a sufficiently large sample to calculate
$\mathbf{\bar{Q}}$ and $\Sigma_Q$ accurately.

%
\begin{figure}
\begin{center}
\resizebox{0.3\textwidth}{!}{%
  \includegraphics{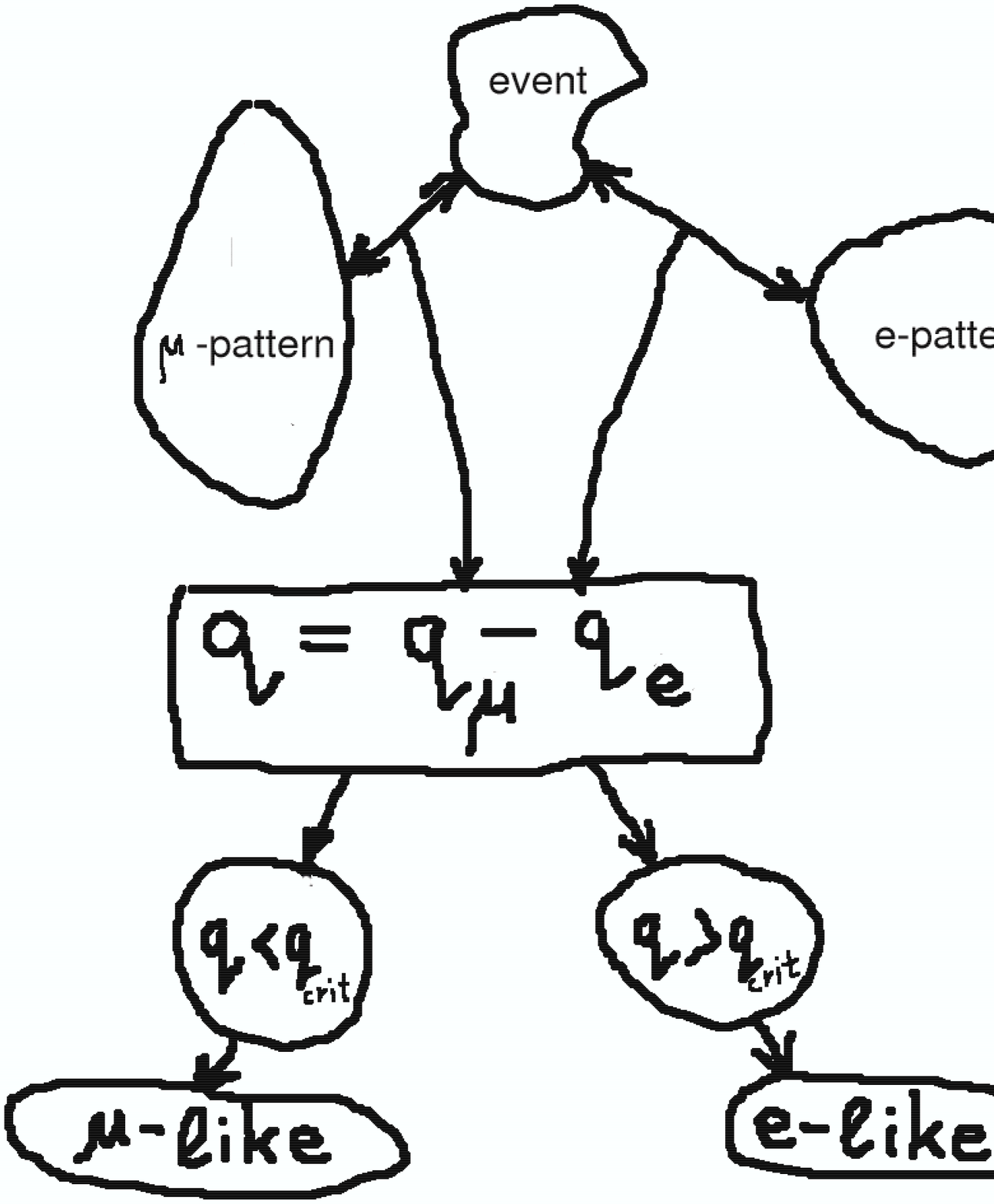}
  }
\end{center}
\caption{ schematic view of pattern recognition.}
\label{fig:1}
\end{figure}
 
Our approach assumes that one can calculate the mean image vector
$\mathbf{\bar{Q}}$ and fluctuation vector 
$\delta \mathbf{Q}$
using the approximations of the Cherenkov light mean angular
 distributions and their fluctuations. Simulation shows that
 correlation coefficients of single PMT contributions amount to
 0.6(0.8) for the neighbouring PMTs and go down to 0.1(0.1) 
for distant PMTs  in electron(muon) events, 
which makes the covariance matrix diagonal-like.

 Thus, we neglect all correlations between $Q_j$ and
the resulting normal distribution becomes simpler: \\
$$
p(\mathbf{Q}; \omega_i,E_0,\mathbf{r_0},
\mathbf{\theta_0}) \,
= \, (2 \pi)^{-N/2} \cdot \left (\prod \limits_{j=1}^N \delta Q_j^
{\omega_i} \right)^{1/2} 
$$
$$
\;\;\;\;\;\;\;
\times exp \left \{ - \sum \limits_{j=1}^N \frac{\left ( Q_j - Q_j^{\omega_i}
 \right )^2}{\delta Q_j^{\omega_i}} \right \} \; , \; \; \; \; (2)
$$

For a certain event geometry (starting point $\mathbf{r_0}$ and
 direction $\mathbf{\theta_0}$
of the particle) and energy $E_0$ we calculate mean pattern images
$Q_j^{\omega_i} \, = \, Q_j^{e,\mu}(E_0,\mathbf{r_0},
\mathbf{\theta_0})  \;$
 and their deviations
$\delta Q_j^{\omega_i} \, = \, \delta Q_j^{e,\mu}(E_0,\mathbf{r_0},
\mathbf{\theta_0})  \;$: \\
\\
$
Q_j^{e,\mu}(E_0,\mathbf{r_0},
\mathbf{\theta_0}) \, 
$
$$
= \, \sum
\limits_{k=1}^n \frac{S}{\rho_{j,k}^2} \cdot cos \chi_{j,k} \cdot
exp \left(-\frac{\rho_{j,k}}{\lambda_{abs}} \right) \cdot
F_k^{e,\mu}(\theta_{j,k}) \; ,
\; \; \; \; \; \; \; \; \; (3)
$$

$$
\delta Q_j^{e,\mu}(E_0,\mathbf{r_0},\mathbf{\theta_0})
 \, = \, \sum
\limits_{k=1}^n \left [\frac{S}{\rho_{j,k}^2} \cdot cos \chi_{j,k} \cdot
exp \left(-\frac{\rho_{j,k}}{\lambda_{abs}} \right) \right ]^2 
$$
$$
\times \left [ F_k^{e,\mu}(\theta_{j,k}) \cdot
\delta_k^{e,\mu}(\theta_{j,k}) \right ]^2 \; ,
\; \; \; \; \; \; \; \; \; (4)
\vspace{0.2cm}
$$
where $k$ is the segment index, $n$ is the number of
segments in a track/shower, $S$ is the PMT area, $\rho_{j,k}$ is the distance
between a segment center
and a PMT center(see Fig.~2), $\chi_{j,k}$ is the cosine of the angle between
$\mathbf{\rho_{j,k}}$ and
a PMT axis (which is normal to the tank surface), $\theta_{j,k}$ is
the radiation angle from the segment
center to the PMT center, and $\lambda_{abs}$ is light absorption 
length in water.
And $F_k^{e,\mu}(\theta_{j,k})$ is the mean angular distribution function,
$\delta_k^{e,\mu}(\theta_{j,k})$ is its relative fluctuation. 

Eq.(4) implies there are no correlations between the contribution 
from different segments of track/shower which is close to 
reality: in simulated electron event samples correlation coefficients 
between neighbouring segments do not exceed ~0.4 and amount to ~0.1 for 
distant segments. In muon events characteristic values of the 
coefficients between different segments (including neighbouring ones) 
are even smaller: ~0.1; the exceptions are the
trailing (last) segments which emit
about 100 times less than the other ones but are 
tightly correlated due to the muon decay process. 

Then we apply Bayes' decision rule which minimizes the error of
misidentification
assuming that prior probabilities of both types of events are equal:
\\
\\
$
r \, = \,
 \frac{P(e/\mathbf{Q})} {P(\mu/\mathbf{Q})} \, 
= \,
\frac{p (\mathbf{Q}/e  )}
{p (\mathbf{Q}/\mu )} =\, 
$
$$
= \,
\frac{\left (\prod \limits_{j=1}^N \delta Q_j^{e} \right )^{1/2}}
{\left (\prod \limits_{j=1}^N \delta Q_j^{\mu} \right )^{1/2}} \cdot
\frac{exp \left \{ - \sum \limits_{j=1}^N \left ( Q_j - Q_j^{e} \right )^2/
\delta Q_j^{e} \right \}}
{exp \left \{ - \sum \limits_{j=1}^N \left ( Q_j - Q_j^{\mu} \right )^2/
\delta Q_j^{\mu} \right \}} \; ,
\; \; \; (5)
$$
where $p \left (\mathbf{Q}/e \right )$ and
$p \left (\mathbf{Q}/\mu \right )$ come from Eq.(2).

The simplest criterion, $q$, which we use to define the type of event is: \\
$$
q \, = \, ln \, r \, = \, q_{\mu} \, - \, q_{el} \, + \, C \;
 , \; \; \; \; \; \; \; \; \;(6)
$$
$$
q_{\mu} \, = \, \sum \limits_{j=1}^N \left ( Q_j^{\, exp} \, - \, Q_j^{\, \mu}
\right )^2 / \delta Q_j^{\, \mu} \; , \; \; \; \; (7-a)
$$
$$
q_{el} \, = \, \sum \limits_{j=1}^N \left ( Q_j^{\, exp} \, - \, Q_j^{\, e}
\right )^2 / \delta Q_j^{\, e}  \; , \; \; \; \; (7-b)
$$
$$
C \, = \, \frac{1}{2} \, ln \, \frac{\prod \limits_{j=1}^N \delta Q_j^{e}}
{\prod \limits_{j=1}^N \delta Q_j^{\mu}}  \; , \; \; \; \; (7-c)
$$
where $Q_j^{\, exp}$ is the contribution to $j$-th PMT in
the ``experimental'' image under consideration.
The event is considered to be $e$-like if $q \, > \, 0$ and $\mu$-like
if  $ q \, < \, 0$.

A slightly more general form of the criterion can help to reduce the errors of
type definition: \\
$$
q \, = \, q_{\mu} \, - \, A \cdot q_{el} \, + \, B \; , \; \; \; \; (8)
$$
where $A,B$ are tuned to minimize the identification errors; the usage
 is the same as above. 
\begin{figure}
\begin{center}
\rotatebox{180}{%
\resizebox{0.3\textwidth}{!}{
  \includegraphics{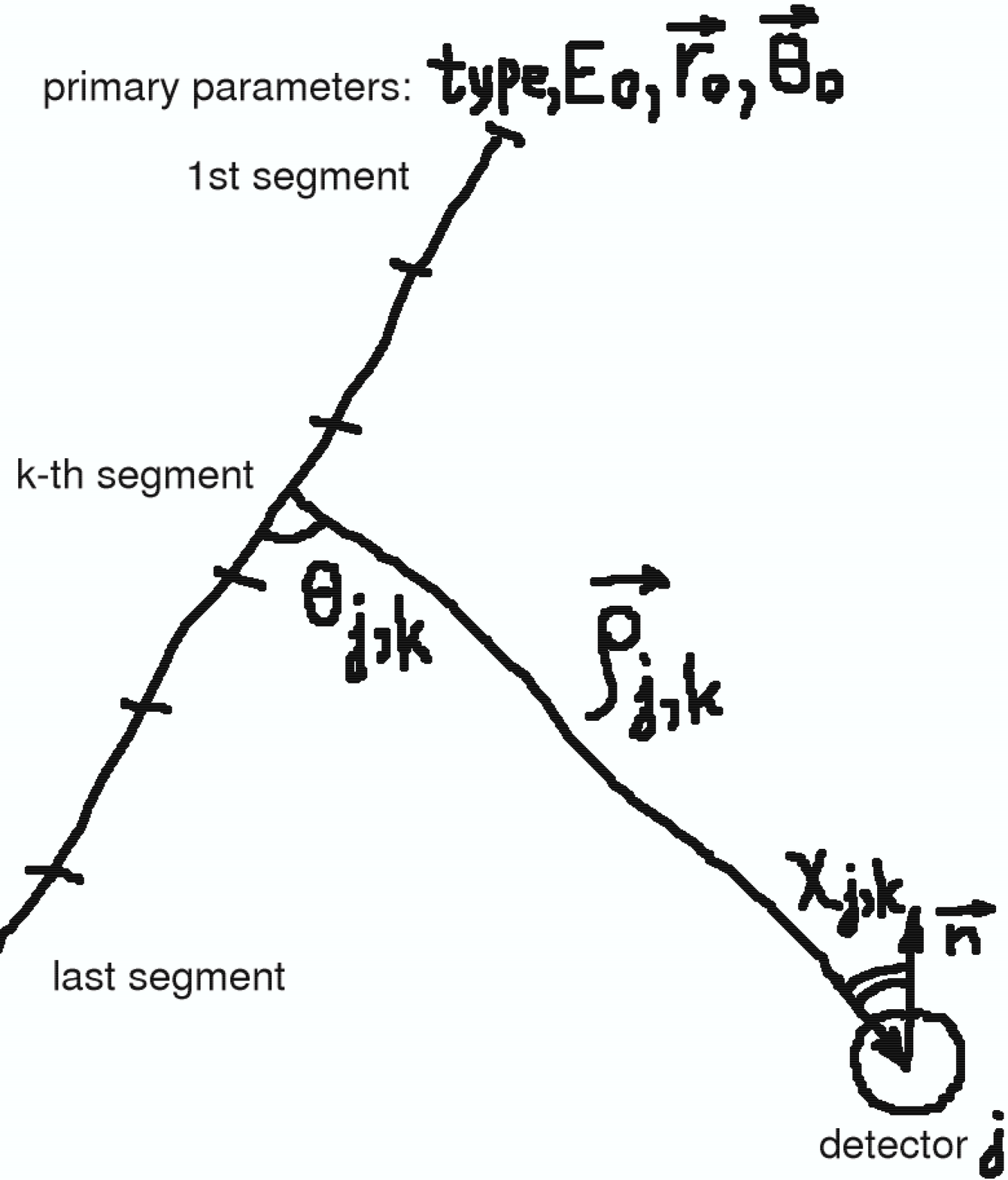}
}}

\end{center} 
\caption{\label{fig:2} geometry for eqs.3 and 4.}
\end{figure}
 

%
%

\section{Examination of the procedure for discriminating
electrons from muons}
\label{sec:3}
%

\subsection{A virtual 1 kilo-ton water tank detector for the KEK experiment}

\label{sec:3.1}
%

  As explained in the preceeding paper\cite{r1}, the SK 
discrimination procedure was constructed on the following 
assumptions in the Kamiokande analysis -- the predecessor to SK :
$$
N_{i,exp}(direct)=\alpha_e \times N_{MC}(\theta_i,p_e) 
\times \left( \frac{16.9}{l_i} \right)^{\gamma} 
$$
$$
\times \mbox{exp} 
\left( -\frac{l_i}{L} \right)
\times f(\Theta),\; \;(9)
$$
\\

$
N_{i,exp}(direct)=
$
$$
=\left\{ \alpha_{\mu} \times \frac{1}{l_i(sin\theta_i+l_i 
\times (\frac{d\theta}{dx}))} \times sin^2 \theta_i + N_{i,knock}(\theta_i) 
\right\} \times 
$$
$$
\times \mbox{exp} \left( -\frac{l_i}{L} \right)
\times f(\Theta),\; \; \; \; \; \; \; \; \;(10)
$$

$$
Prob=\frac{1}{\sqrt{2\pi}\sigma}\mbox{exp} \left( -\frac{(N_{obs}-N_{exp})^2}
{2\sigma^2} \right),\; \; \; \; \; \; \; \;(11)
$$

$
P_{pattern}(e)  = 
$
$$
=\displaystyle \mbox{exp} \left\{ -\frac{1}{2}\left( 
 \frac{\chi^2(e)-min[\chi^2(e),
\chi^2(\mu)]}{\sigma_{\chi^2}} \right)^2 \right\}, \; \;(12-a)\\
$$
$
P_{pattern}(\mu)  = 
$
$$
=\displaystyle \mbox{exp} \left\{ -\frac{1}{2}\left( 
 \frac{\chi^2(\mu)-min[\chi^2(e),
\chi^2(\mu)]}{\sigma_{\chi^2}} \right)^2 \right\}. \; \;(12-b)\\
$$

$
P_{angle}(e)=
$
$$
  =  constant \times \displaystyle \mbox{exp} \left\{ -\frac{1}{2} 
\left( \frac{\theta_{exp}(e)-\theta_{obs}}{\Delta \theta} \right)^2 \right\}
,\; \; \; \;(13-a) \\
$$
$
P_{angle}(\mu)= 
$
$$
 =  constant \times \displaystyle \mbox{exp} \left\{ -\frac{1}{2} 
\left( \frac{\theta_{exp}(\mu)-\theta_{obs}}{\Delta \theta} \right)^2 \right\}
,\; \; \; \;(13-b)
$$


$$
P(e)  =  P_{pattern}(e) \times P_{angle}(e), \; \; \; \; \; \; \; \; (14-a) \\
$$
$$
P(\mu)  =  P_{pattern}(\mu) \times P_{angle}(\mu).\; \;  \; \; \; \; \; \;(14-b)
$$

  These expressions was given firstly in Takita \cite{r2Takita}.\\  

However, in order to show the validity of this discrimination procedure which 
had never been confirmed by accerelator
 beam, the SK group performed an accelerator experiment using the
KEK 12~GeV proton synchrotron (Kasuga\cite{r3Kasuga}, Kasuga et al.\cite{r4Kasuga}, Sakai\cite{r5Sakai},
 Kasuga\cite{r6Kasuga2}). For the confirmation, one kiloton water 
 Cherenkov detector 
 was constructed and the SK group studied the
particle identification capabilities over the momentum range 250~MeV/c
to 1000~MeV/c for muons and 100~MeV/c to 1000~MeV for electrons.  They
concluded that the particle misidentification probabilities  
with SK discrimination procedure
is less than 1 to 5\% (Sakai, p.141 \cite{r5Sakai}, Kasuga, 
p.77 \cite{r6Kasuga2}).  Based on these procedures, the SK
group have reported experimental results which have lead to the
claimed existence of neutrino oscillations.  To study the positional
dependence of the discrimination, data was taken for 10 injection
points in the 1-kiloton detector: WUS, ESS, WUC, WUD, ... (See
Sakai\cite{r5Sakai} in detail).  We have constructed a virtual 1~kilo-ton
water Cherenkov detector, including the same PMT configuration, as the
SK test one in our computer.  In the virtual detector we simulate
physical events and examine the influence of the virtual physical
events on the virtual detector.

In 
Figure~3 we give the relation between the 
emitted Cherenkov light
and the detected Cherenkov light for the WDS injection point 
(see Sakai's paper).  The emitted Cherenkov light denotes the total
Cherenkov light produced by the charged particle concerned, while the
detected light records only that which is measured by the detector.
This relation holds almost exactly, irrespective of the
generation point, because the absorption length is so large compared
with the dimensions of the 1~kilo-ton detector.
\begin{figure}

\begin{center}
\resizebox{0.35\textwidth}{!}{%
  \includegraphics{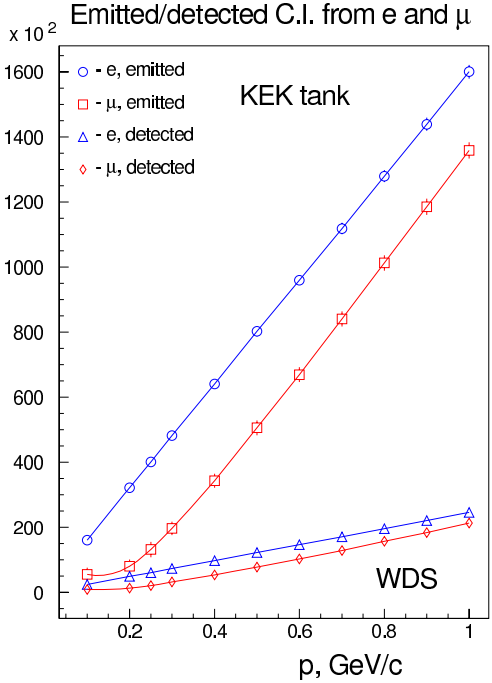}
  }
\end{center}
\caption{\label{fig:3} The average behavior of the Cherenkov
light for both emitted and detected as the functions of 
momentum for muons and electrons which are generated at WDS.}
\end{figure}

It is clear from Figure 3 that there is big difference as 
for the Cherenkov light quantity between the emitted Cherenkov 
light and the detected one. The emitted Cherenkov light denotes, 
by the definition , the total Che-\\
renkov light, while the detected 
Cherenkov light denotes only small part of the emitted one. 
Consequently, we expect variety of the patterns of the detected 
Cherenkov lights for the events concerned due to the PMT 
configuration. 

Using the discrimination procedure developed in section~2 
(Eq.(8), $A$=1, $B$=$q_{crit}$), we try to
discriminate 300~MeV electrons from 500~MeV muons, both of which
result, on average, in the production of the same amount of Cherenkov
light.  The result is given in 
Figure~4. 
As can be seen,
discriminating between the electron and muon seems to be very
difficult.
In Table~1 we give misidentification probability for different
$q_{crit}$.  The parameter $q_{crit}$ denotes a criterion for
discriminating between muons and electrons: $e \, \rightarrow \, \mu$
denotes the probability for an electron to be misidentified as a muon,
and $\mu \, \rightarrow \, e$ denotes the probability for a muon to be
misidentified as an electron.  It is easily understood from
Figure~4 
that a non-negligible number of muon events penetrate into
the electron events region, so that some muon events may be
misidentified as electrons.

\begin{center}
\begin{table}
\caption{\label{Table:1} 
 Misidentification probability between 300MeV electron events 
and 500MeV muon  events in the virtual KEK 1 kiloton tank
 detector, taking into account forward scattering only.
 }
\vspace{0.5cm}
\hspace*{1cm}
\begin{tabular}{|c|c|c|c|}

\hline
&&&\\
$run$&$q_{crit}$&$e{\rightarrow}{\mu}{\:}{\%}$&${\mu}{\rightarrow}e{\:}{\%}$\\
&&&\\
\hline
 1  &  -4.00e2  &  1.67  &  25.67\\
 2  &  -5.00e2  &  2.67  &  22.33\\
 3  &  -6.00e2  &  6.67  &  19.67\\
 4  &  -7.00e2  & 12.67  &  18.67\\
 5  &  -8.00e2  & 20.33  &  17.30\\
 6  &  -9.10e2  & 27.33  &  15.67\\
 7  &  -1.00e3  & 35.67  &  12.67\\
\hline

\end{tabular}
\end{table}
\end{center}

\begin{figure}
\vspace{0.5cm}
\begin{center}
\resizebox{0.3\textwidth}{!}{%
  \includegraphics{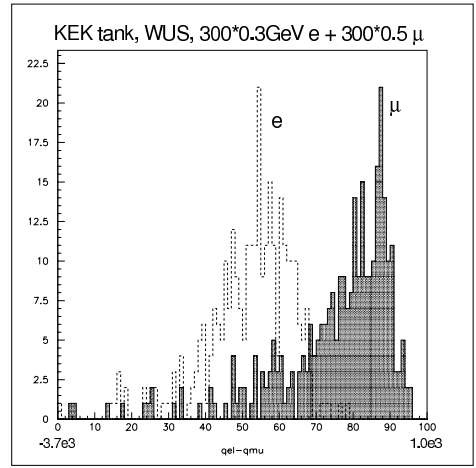}
  }
\vspace{0.5cm}
\caption{\label{fig:4} The $q_{el}$-$q_{\mu}$ distribution for 
simulated events for 300 MeV electrons and 500 Mev muons in 
the KEK kiloton tank with angular distribution function
 describing only the forward hemisphere. 
These particles produce approximately the same amount of
 Cherenkov light. 
  The notation " 300*0.3 GeV e " denotes the reslutls for 300 
MeV electrons from 300 simulations.}
\end{center}
\end{figure}

Thus, we carefully examined the muon events which were located in the
electron event parameter space, i.e., where electron events are found
with high probabilities. As a result, we found that the majority of
such muon events are $\mu-e$ decay events which should be identified
as muon events.
 However, from the view point of the pattern
recognition, we could not recognize whether the events in the gray
zone between the `clearly electron events' and the `clearly muon
events' were electrons or muons. To resolve this, we extend our
angular distribution function, which was defined only within the
forward hemisphere, to include the backward hemisphere and 
redefine $q_{el}$ and $q_{\mu}$ in accordance with this extension.

After such a manipulation,
one gets much smaller contamination of electron domain by muon events.

In Figure~5, we give the
revised $q_{el}$-$q_{\mu}$ distribution where there is almost no 
intersection between the muon events and electron events. 
 In Table~2, we give the
misidentification probability based on the revised procedure.
Comparing Table~2 with Table~1, we find that misidentification
probability has decreased remarkably.

\begin{center}
\begin{table}
\caption{\label{Table:2} 
Misidentification probability between 300MeV electron events and 500MeV muon
 events in the virtual KEK 1 kiloton tank detector, taking into account
  forward and backward scattering.
 }
\vspace{5mm}
\hspace*{15mm}
\begin{tabular}{|c|c|c|c|}

\hline
&&&\\
$run$&$q_{crit}$&$e{\rightarrow}{\mu}{\:}{\%}$&${\mu}{\rightarrow}e{\:}{\%}$\\
&&&\\
\hline
1   &    2.00e2   &         0.00    &       4.67  \\
2   &    1.00e2   &         0.33    &       3.67  \\
3   &    0.00     &         0.67    &       3.00  \\
4   &   -1.00e2   &         0.67    &       2.00  \\
5   &   -2.00e2   &         2.00    &       2.00  \\
6   &   -3.10e2   &         3.67    &       2.00  \\
7   &   -4.00e2   &        12.00    &       2.00  \\
8   &   -5.00e2   &        20.33    &       1.67  \\
9   &   -6.00e2   &        30.00    &       1.67  \\
10  &   -7.00e2   &        43.67    &       1.33  \\
\hline

\end{tabular}
\end{table}
\end{center}

\begin{figure}
\vspace{0.5cm}
\begin{center}
\resizebox{0.3\textwidth}{!}{%
  \includegraphics{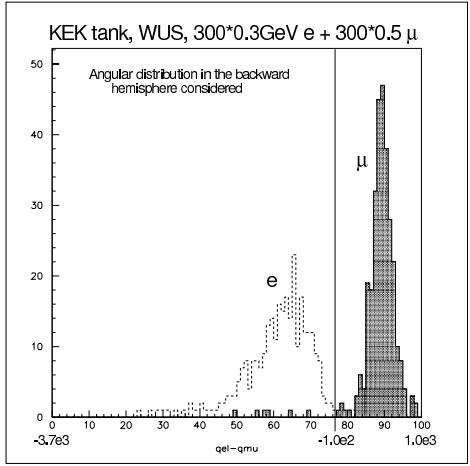}
  }
\vspace{0.5cm}
\caption{\label{fig:5} The $q_{el}$-$q_{\mu}$ distribution for
 simulated events for 300 MeV electrons and 500 Mev muons in 
the KEK kiloton tank with angular distribution function
 describing both forward and backward hemisphere. 
These particles produce approximately the same amount of
 Cherenkov light. 
 The notations are the same as for Figure~4.}
\end{center}
\end{figure}

The probability distribution of the standard SK estimator for particle
identification, if it were assumed to be remotely comparable, should
be compared to Figure~4, not Figure~5, because 
the effect of
$\mu-e$ decay was not considered correctly in the SK simulation (See
Sakai,p.154, \cite{r5Sakai}).  There are, thus, big differences between
our result and the SK results (For example see Figure~3 in
Sakai\cite{r5Sakai}).  Namely, our initial result 
(Figure~4) could not
reliably separate electron events from muon events, although it is
claimed the standard SK analysis separates electron events from muon
events almost completely.  We can not imagine such a complete
discrimination could be realized by the SK procedure.

The big difference between the ``complete'' discrimination by the SK
(Figure~3 in Kasuga et al.\cite{r4Kasuga}) and the poorer
discrimination obtained by us (Figure~4) extends further, if we
take into consideration the uncertainties added by the real detector
to our Figure~4. This is because we do not consider the effect of
photoelectrons produced from the parent Cherenkov light, which obeys a
stochastic physical process, or the effect of scattered Cherenkov
light.  These two factors must increase the uncertainties in the
discrimination between muons and electrons.


\subsection{Discrimination of electrons from muons in
 the Super-Kamiokande detector in our simulation}
\label{sec:3.2}

Also, we have constructed a virtual SK detector incorporating the
actual PMT configuration and have carried out the experiment
numerically in our virtual SK detector. Based on the revised procedure 
for the discrimination of electron events from muon events developed 
in the previous section, we have discriminated electron events from muon
events.

\begin{figure}
\begin{center}
\resizebox{0.4\textwidth}{!}{%
  \includegraphics{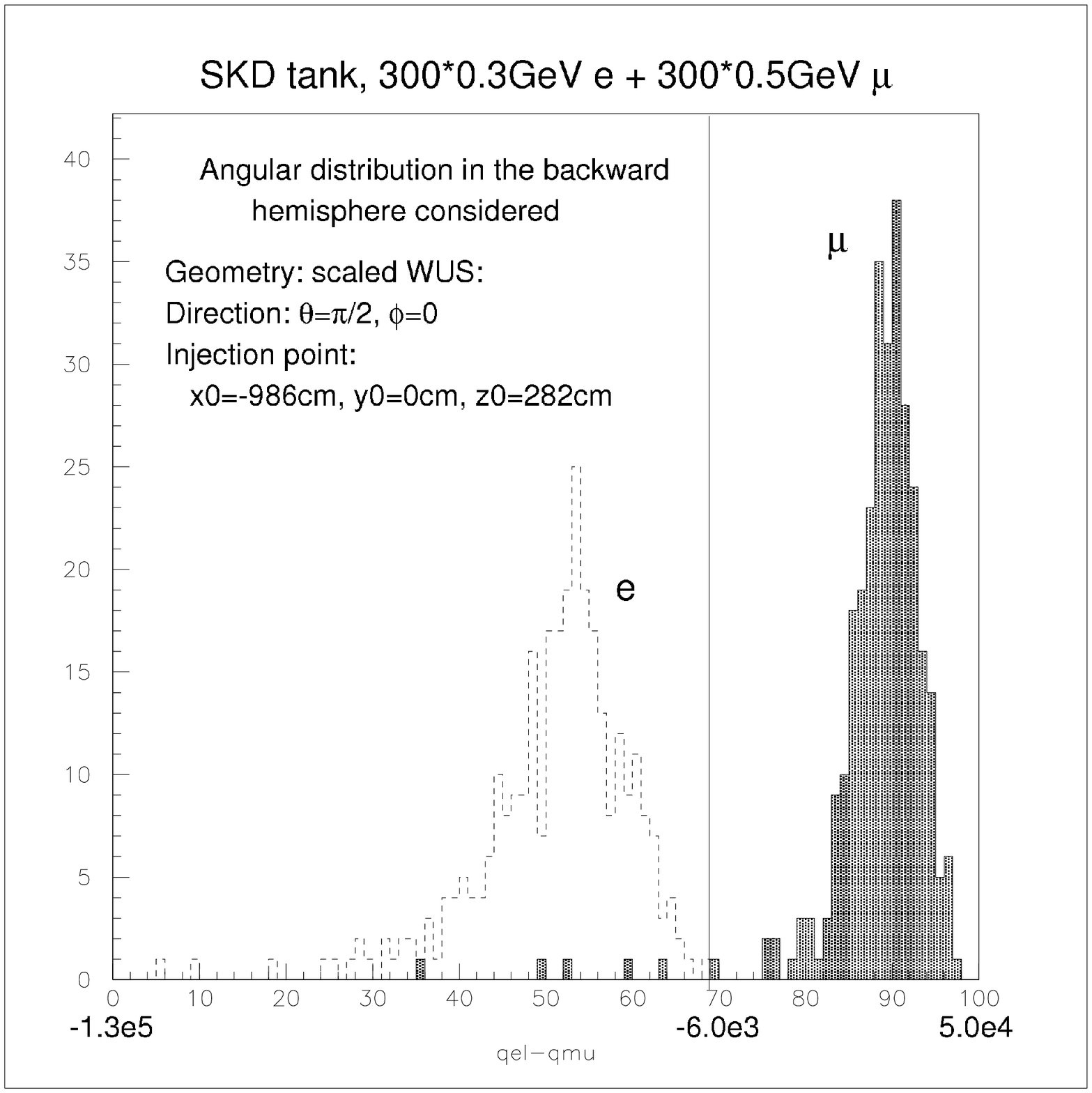}
}
\vspace{0.5cm}
\caption{\label{fig:6} The $q_{el}$-$q_{\mu}$ distribution for 
simulated events for
 300 MeV electrons and 500 MeV muons in the virtual SK detector. 
The notations are the same as for Figure~4.}

\vspace{10mm}
\hspace{30mm}
\resizebox{0.4\textwidth}{!}{%
  \includegraphics{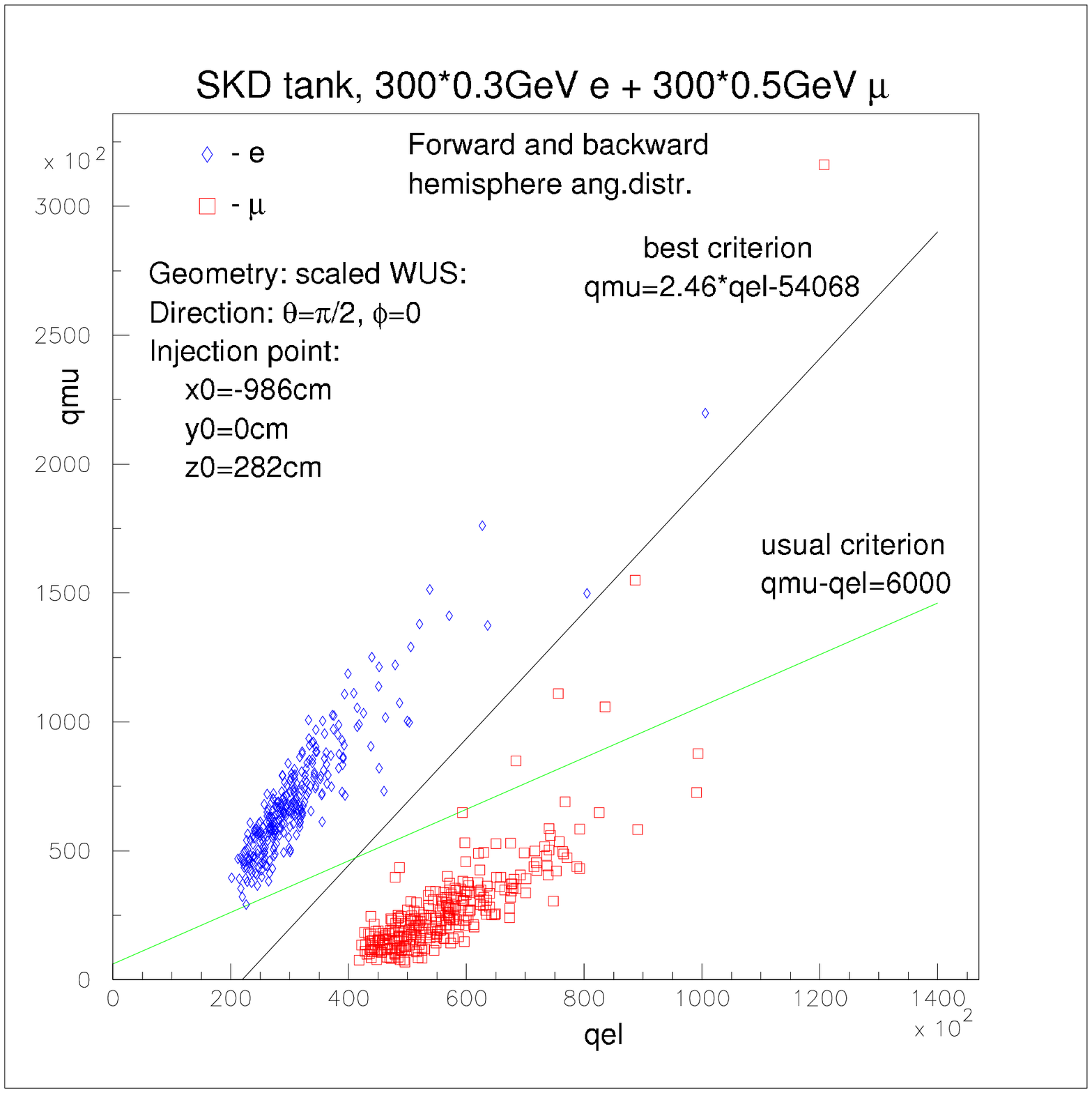}
}
\vspace{0.5cm}
\caption{\label{fig:7} The scatter plot for $q_{el}$ and $q_{\mu}$ for 
300 MeV electrons
 and 500 MeV muons considered in Figure~6.} 
\end{center}
\end{figure}

In Figure~6 we give the discrimination of the muon 
events from
electron events in the virtual SK detector by the revised method. We conclude
that we could discriminate muon events from electron events using the
revised method with high probability. 
In Figure~7 
we show the scatter plot between $q_{e}$ and $q_{\mu}$. From 
Figure~6 and Figure~7, 
we can conclude that our procedure for the discrimination
of electron events from muon events works well.

\begin{figure}
\begin{center}
\resizebox{0.4\textwidth}{!}{%
  \includegraphics{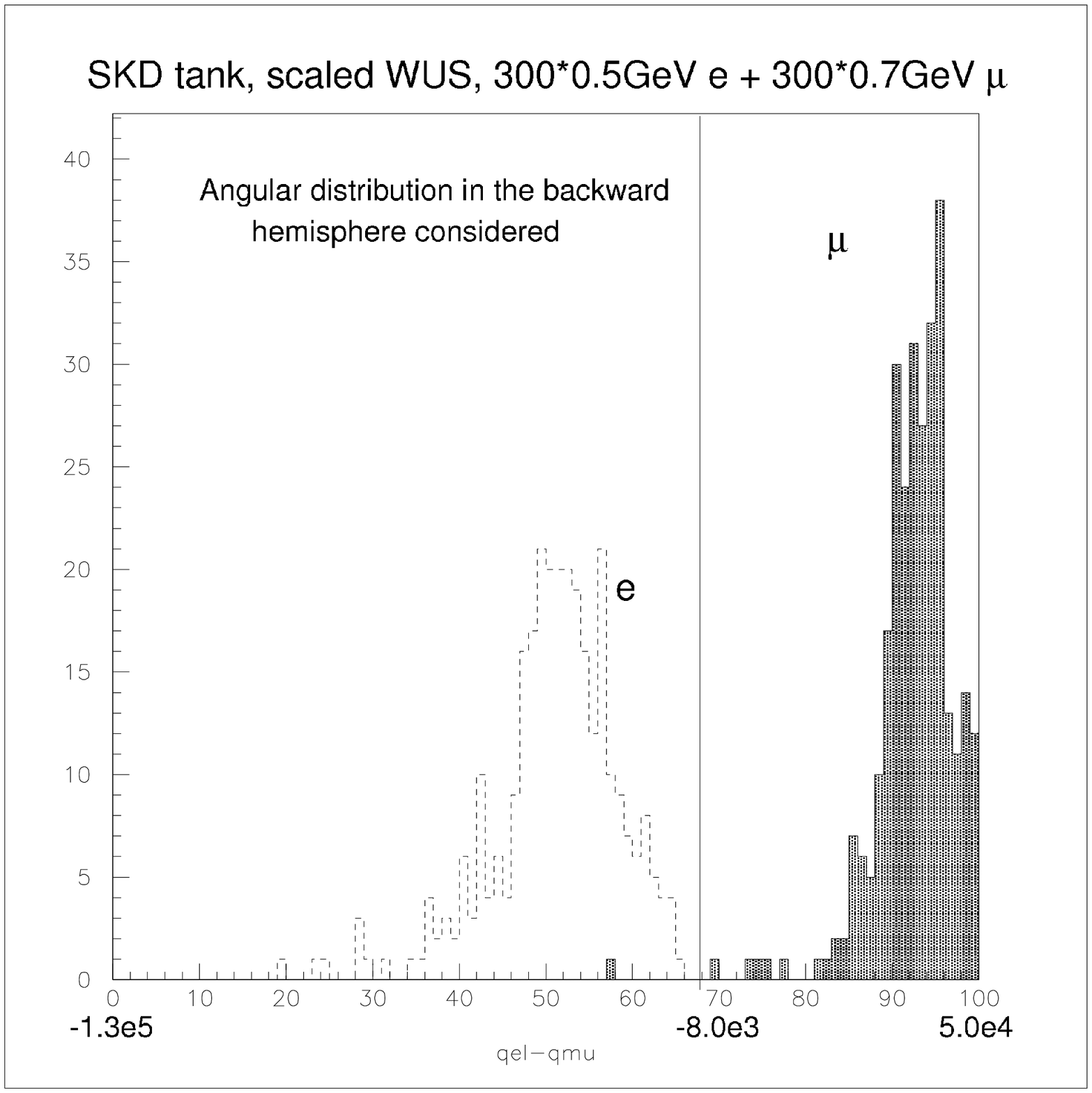}
}
\vspace{0.5cm}
\caption{\label{fig:8} The $q_{el}$-$q_{\mu}$ distribution for 
simulated events for
 500 MeV electrons and 700 MeV muons in the virtual SK detector. 
The notations are the same as for Figure~4.}
\vspace{10mm}
\hspace{30mm}
\resizebox{0.4\textwidth}{!}{%
  \includegraphics{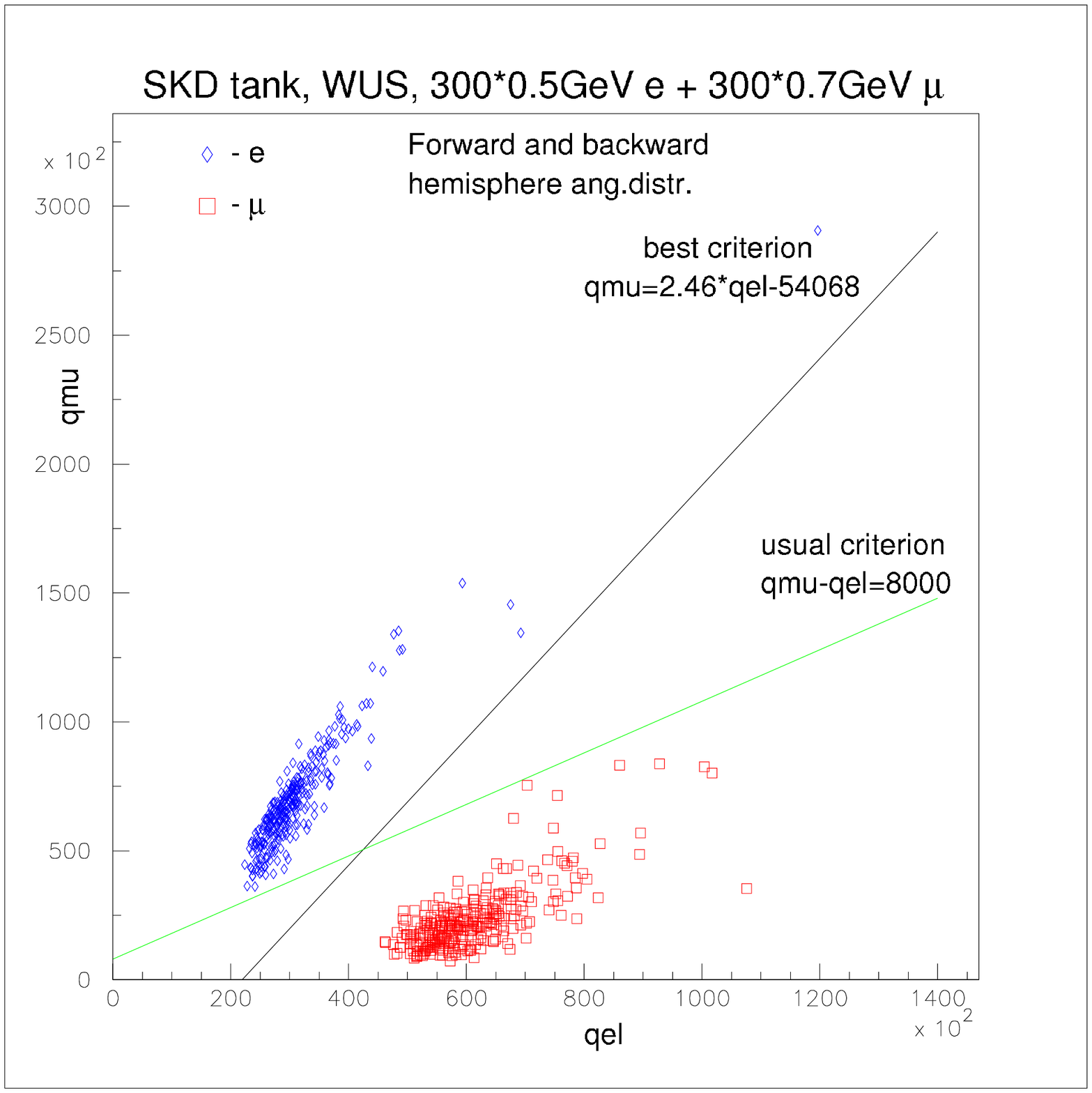}
}
\vspace{0.5cm}
\caption{\label{fig:9} The scatter plot for $q_{el}$ and $q_{\mu}$ for 
500 MeV electrons 
and 700 MeV muons considered in Figure~8.} 
\end{center}
\end{figure}

In Figure~8 and Figure~10 we show the discrimination 
of muon events from electron events for a comparison of 500~MeV electrons 
with 700~MeV muons, and 700~MeV electrons with 900~MeV muons, respectively.
These electron and muon energies were chosen as they yield roughly the
same amount of Cherenkov light.

\begin{figure}
\begin{center}
\resizebox{0.4\textwidth}{!}{%
  \includegraphics{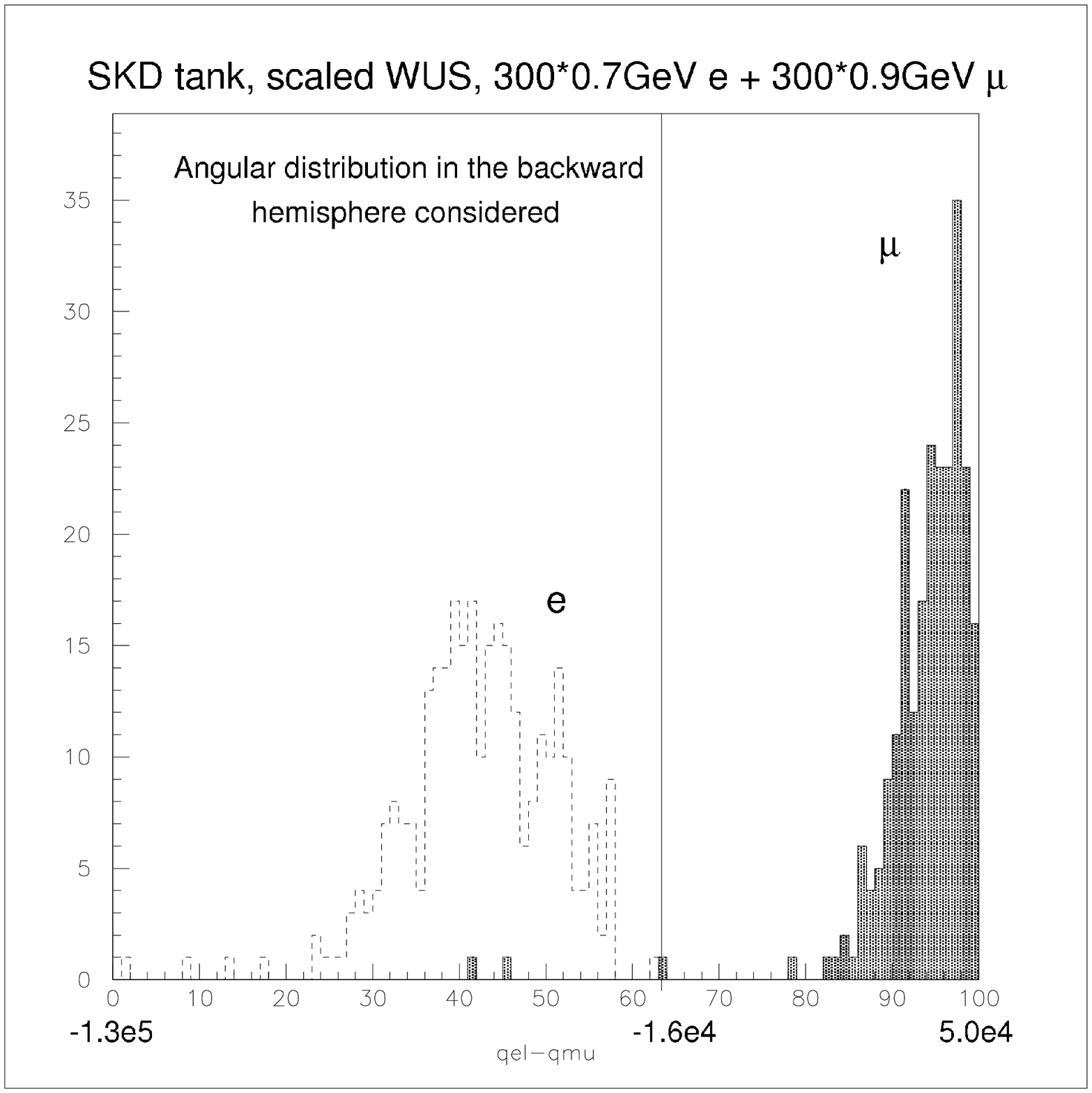}
}
\vspace{0.5cm}
\caption{\label{fig:10} The $q_{el}$-$q_{\mu}$ distribution for 
simulated events for
 700 MeV electrons and 900 MeV muons in the virtual SK detector. The 
notations are the same as for Figure~4.}
\vspace{1cm}
\resizebox{0.4\textwidth}{!}{%
  \includegraphics{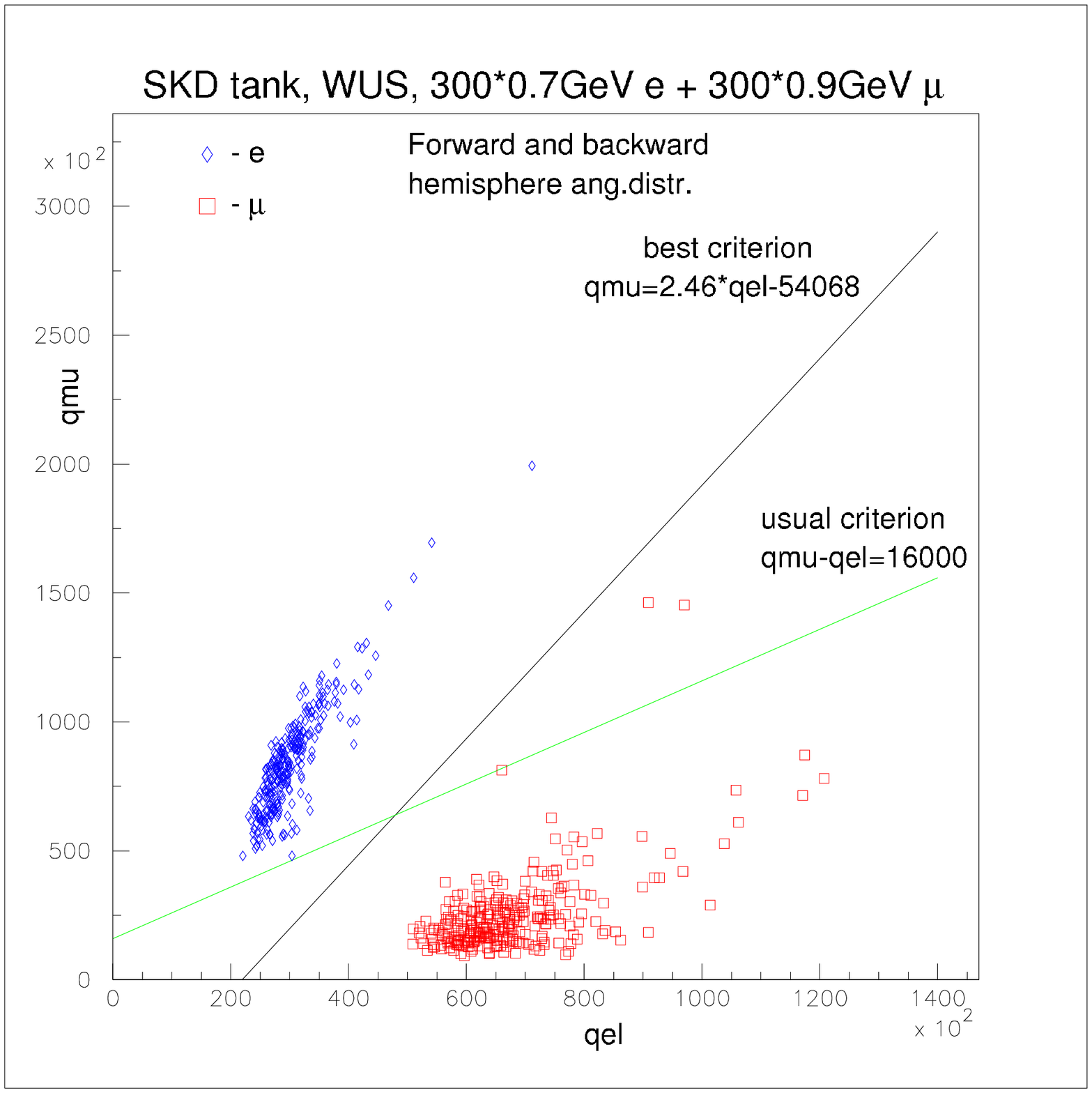}
}
\vspace{0.5cm}
\caption{\label{fig:11} The scatter plot for $q_{el}$ and $q_{\mu}$ for 
700 MeV electrons
 and 900 MeV muons considered in Figure~10.} 
\end{center}
\end{figure}

 Also, scatter plots corresponding to
Figure~8 and Figure~10 are given in Figure~9 
and Figure~11, respectively.  
Comparing Figure~6 with Figure~8,
 it is apparent
that the latter allows a better discrimination. Comparing 
Figure~8
with Figure~10, the latter is also better than the former for
discrimination.  The reason for this is that the generation of
Cherenkov light for muon events becomes more different from that of
the electron as their primary energies increase.  Discrimination
procedures of muon events from electron events in the SK detector are
examined for events whose primary particles start at the ``scaled
WUS'' position ($x_0$=986~cm, $y_0$=0~cm, $z_0$=282~cm, $\theta 
\, = \, \pi/2$and $\varphi \, = \, 0$).

%

\section{Error Distribution for the vertex point and direction
 for muon and electron events}
\label{sec:4}

\subsection{TDC procedure}
\label{sec:4.1}

The SK analysis introduces a ``TDC procedure'' to determine the vertex
position.  The principle of the TDC procedure is to find the position
where the time residuals, $t_i$, for the PMTs being fitted are
minimized.  The time residual $t_i$ of the i-th PMT is defined as
$$
t_i \, = \, t_i^0 \, - \, \frac{n}{c} \times
\sqrt{(x-x_i)^2 \, + \, (y-y_i)^2 \, + \, (z-z_i)^2}
\; , \; \; \; (15)      
$$
where $t_i^0$ is the hit time of the $i$-th PMT, $(x_i,y_i,z_i)$ is
the position of the $i$-th PMT, $(x,y,z)$ is the effective emitting
point position and $c/n$ is the velocity of the Cherenkov light in
water.  That all the light is emitted from the same {\it effective}
point in space and comes to $j$-th PMT {\it exactly} at the moment
corresponding to the mean time $\bar{t_j}$ of the $j$-th PMT Cherenkov
pulse is not really true. A simple equation for time residual used in
a $\chi^2$-like sum (the system of linear equations for the effective
point coordinates $(x^*,y^*,z^*)$ is overdefined!) can give effective
point estimate after the sum's minimization (with respect to
$(x^*,y^*,z^*)$) even if the original assumption is not valid. The
effective point thus deduced does not coincide with the center-of mass
of the light emitting system ($e$-shower or $\mu$-track) because of
specific mechanism of Cherenkov pulse formation and will usually
differ from the event starting (injection) point.

The TDC procedure based only on $\bar{t_j}$ cannot estimate the event
direction because to define a direction one needs at least two
points. Direction estimates could be obtained as a result of Cherenkov
pulse shape analysis for each sufficiently illuminated optical module
if the PMT and electronics are fast enough for such analysis.

Sakai shows the time residual distribution of typical event
 (a 1~GeV/c, electron) which is distributed over 50 
nanoseconds (Sakai, p.38 \cite{r5Sakai}), assuming
 a point-like source. We simulate
the Cherenkov light in the cascade shower using GEANT~3.21 and the
tools we have developed.  In 
Figure~12
we give one example for the
time residual distribution for a 1~GeV primary electron based on
Eq.(15) with the use of the detailed simulation of 
the cascade shower.
In our calculation, we simulate shower particles 
and the accompanying
Cherenkov light due to shower particle concerned in an exact
way. Then, we know the starting point of the primary electron.
Shifting the starting point from the real point to a range of
artificial ones, we can obtain the time residual 
distribution for each
position, and examples are given in Figure~12.\\

\begin{figure*}
\hspace*{1cm}
\rotatebox{90}{%
\resizebox{0.5\textwidth}{!}{
  \includegraphics{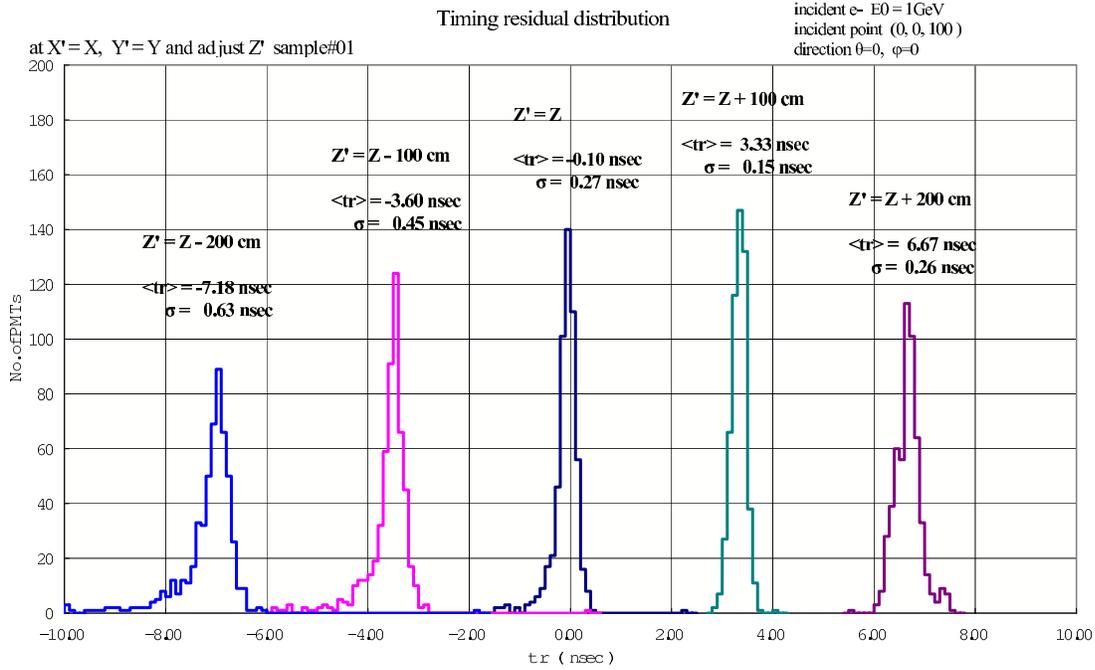}
}}

\caption{\label{fig:12}  One example of the time residual distributions for 
1 GeV electron primary shower, assuming different vertex positions.}
\end{figure*}

Among the five
different starting points, which includes the true one, the smallest
standard deviation is obtained in the case of $ Z'=Z+100 $~(cm), where
$Z'$ and $Z$ denote the assumed vertex point and the real vertex
point, respectively. Thus, the apparently most probable vertex point
is not real one, but is offset by 100~cm from the real one.  Of
course, this is only one example and not average behavior.  However we
examined many individual cases and confirm that this is usual
character which should hold even for the average behavior.

The comparison of our simulations with the experimental data from the
SK experiment (Sakai) reveals a large difference.  The width of our
time residuals distribution is within one nanosecond, while the width
for SK is $\sim$50~nanoseconds.  The probable reasons are that we have
not considered the PMT and electronics response functions, and that we
have neglected light scattering.

We calculate the time residuals for electrons of 500~MeV, 1~GeV, 3~GeV
and 5~GeV, assuming that all Cherenkov light comes from certain points
of a shower/track. From these calculations one can see that the
smallest time residuals do not give the vertex position but yield
points shifted from the vertex point along the direction of the
cascade shower, namely, 50 to 100~cm for 500~MeV electrons, 100 to
150~cm for 1~GeV
(Figure\ref{fig:12})
 and 3~GeV electrons, and 150 to 200~cm
for 5~GeV electrons.  Such a tendency is quite understandable if we
consider the size of the shower/track: the effective point should not
be too close to the starting or ending points.  The error of effective
point location by minimizing the width of the distribution amounts to
about 50~cm.

    From the much larger width of the SK time residual distribution it
is clear that in experimental conditions the effective point location
error should be a few times greater because the minimum of the width
as a function of effective point position would be much less
pronounced.

For reasons mentioned above, we conclude that the SK TDC procedure is
not suitable for the determination of an accurate vertex position for
electron events.  The situation for the muon events is essentially the
same but must be worse than in the case of electron events as muon
events have a longer extent than the corresponding electron events.


\subsection{ADC procedure}
\label{sec:4.2}
The standard SK analysis uses an ``ADC procedure'' for event geometry
reconstruction which is based only on the total amount of Cherenkov
light detected by the PMTs.

We have constructed and used an ADC procedure which is based on
the more detailed model for the angular distribution for the Cherenkov light
described above.


\subsubsection{Fundamental procedure for the error distribution for muon
events and electron events: event geometry definition procedure}
\label{sec:4.2.1}
Our geometry reconstruction procedure assumes the type and energy of
event are known. Thus, we only need to determine the event starting point
$\mathbf{r_0}$ and the direction of the primary particle
$\mathbf{\theta_0}$.

\begin{figure}
\begin{center}
\resizebox{0.3\textwidth}{!}{%
  \includegraphics{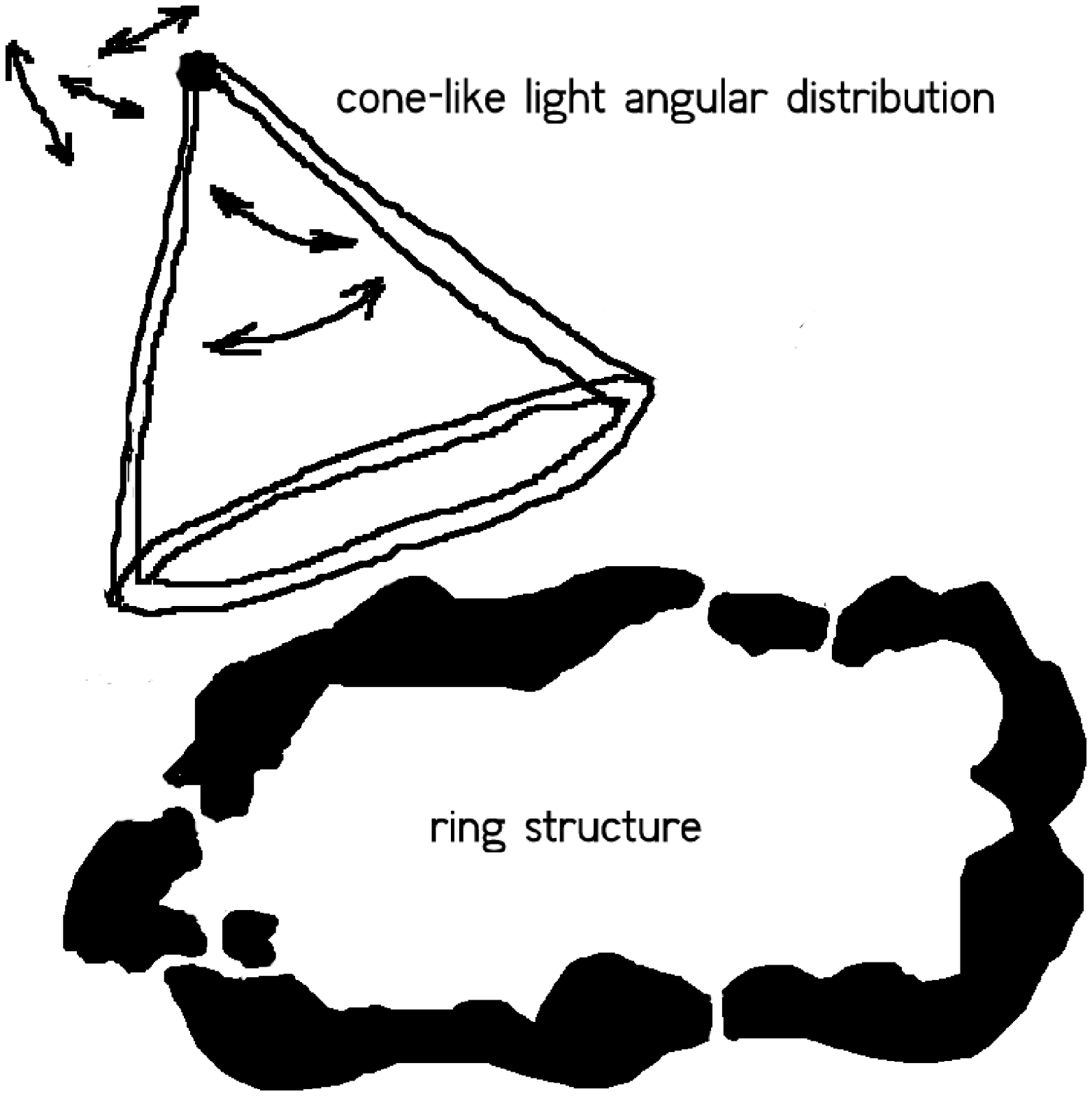}
}

\caption{\label{fig:13} 'cone' figure.}
\end{center}
\end{figure}

\begin{figure}
\begin{center}
\resizebox{0.3\textwidth}{!}{%
  \includegraphics{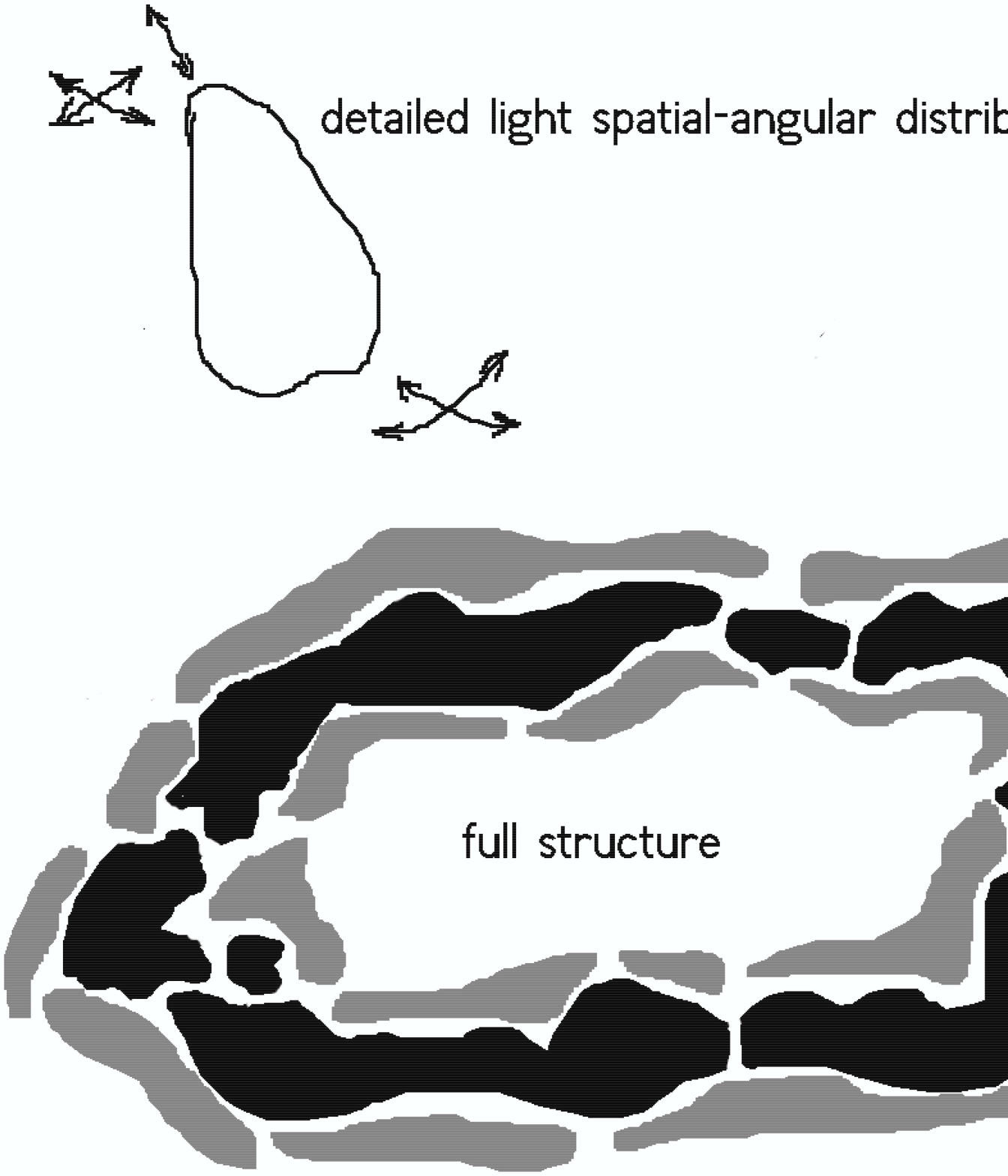}
}

\vspace{-0mm}
\caption{ \label{fig:14} 'full' figure.}
\end{center}
\end{figure}
 

The first step is to find a ring-like (or arc-like) structure
in the image. In muon events we observe two ring-like 
structures because the muon-decay electron produces a 
ring-like or a spot-like structure of its own, however, we 
do not consider muon energies below 500~MeV, so in muon events
we choose the most intensive ring-like structure. This task is
accomplished by row-by-row (or column-by-column) scanning of
the arrays, representing an image, for maxima above a certain
threshold.

The second step is defining the first approximation 
$\; \mathbf{r_1},\mathbf{\theta_1} \;$ 
of geometrical parameters by fitting a ring structure with the
following simple (cone-like) model: all light $\; Q_{tot} \;$
emitted by a $\mu$-track/$e$-shower is assumed to come from
one effective point $W$ on the track with the angular
distribution of light being Heaviside-like: \\


\hspace*{5mm}
$
F(\theta) \, = \, \left \{ 
\begin{array}{cccc}
& 0, &  \theta \, < \bar{\theta} - \Delta \theta   & \\
& a, &  |\theta - \bar{\theta}| \le \Delta \theta  & \\
& 0, &  \theta \, > \bar{\theta} + \Delta \theta  & 
\end{array} \right.,
$
$$
 a:\; \; 2\pi \, \int \limits_0^{\pi} F(\theta)
 \, sin \theta d\theta \, = \, Q_{tot} \; \; \; \; (16)
$$


The zero approximation for geometry parameters could be arbitrary because
the ring-like structure is usually very pronounced and we use the 
cone-like model of Cherenkov light angular distribution with abrupt 
edges, the function to minimize with respect to
$\mathbf{r},\mathbf{\theta}$ is: \\
$$
G(\mathbf{r},\mathbf{\theta}) \, = \,
\sum \limits_{l=1}^M \frac{\left ( Q_l^{\, exp}(\mathbf{r_0},
\mathbf{\theta_0}) \, - \, Q_l(\mathbf{r},
\mathbf{\theta}) \right )^2}{Q_l^{\, exp}(\mathbf{r_0},
\mathbf{\theta_0})}
\; , \; \; (17)
$$
where $l$ is the PMT index within the ring structure, $M$ is the number
of PMTs in the ring structure, $Q_l^{\,
exp}(\mathbf{r_0},\mathbf{\theta_0})$ is the
contribution to $l$th PMT from the event under consideration,
$Q_l(\mathbf{r},\mathbf{\theta})$ is an estimate for
this contribution calculated with the cone-like angular
distribution model described above, $F(\theta)$: \\
$$
Q_l(\mathbf{r},\mathbf{\theta}) \, = \,
 \frac{S}{\rho_{l,W}^2} \cdot cos \chi_{l,W} \cdot
exp \left(-\frac{\rho_{l,W}}{\lambda_{abs}} \right) \cdot
F(\theta_{l,W}) \; , \; \; (18)
$$

The third and final step is the improvement of geometry parameter
estimates by fitting all of the image above a certain threshold, 
$Q_{thr}$, with the pattern image \\
$\left \{ Q_j(E_0,\mathbf{r},\mathbf{\theta}),
\delta Q_j(E_0,\mathbf{r},\mathbf{\theta}) \right \}$
calculated using the detailed model of Cherenkov light angular
distribution. The previous step fit result $\;
\mathbf{r_1},\mathbf{\theta_1} \;$ is used as the first
approximation, and the function to minimize with respect to
$\mathbf{r},\mathbf{\theta}$ is: \\

$
H(\mathbf{r},\mathbf{\theta}) =\, 
$
$$
= \sum \limits_{j: \, Q_j^{\, exp} \ge Q_{thr}}
\frac{\left ( Q_j^{\, exp}(E_0,\mathbf{r_0},
\mathbf{\theta_0}) \, - \, Q_j(E_0,\mathbf{r},
\mathbf{\theta}) \right )^2}
{\delta Q_j(E_0,\mathbf{r},\mathbf{\theta})} \; \; \; (19)
$$

The PMT contribution threshold,  $Q_{thr}$, could be optimized with
respect to the resulting geometry parameter resolution, as we show
later.  The optimal threshold, $Q_{thr, \, opt}$, increases with
primary energy, which could be used to enhance the resolution,
because an estimate of the primary energy can be obtained from the
total amount of Cherenkov light detected quite independently of the
ADC procedure.

Figure~13 shows the process of geometry reconstruction using ring 
structure. Assuming geometry parameters of an event to be 
$\mathbf{r},\mathbf{\theta}$, 
cone-like angular distribution of light (Eq.(16)) is used to evaluate 
contributions (18) to all detectors (PMTs) of the ring structure in 
order to calculate 
$G(\mathbf{r},\mathbf{\theta})$ (Eq.(17))
 which compares this rough light 
pattern with the light ring structure. Adjusting 
$\mathbf{r}$ and $\mathbf{\theta}$
(see arrows in the figure) so that $G$ approaches its minimum one 
obtains a rough estimate of the true geometry parameters 
$\mathbf{r}_0,\mathbf{\theta}_0$.

Figure~14  shows the process of geometry reconstruction 
using full light image of an event. Assuming geometry parameters of 
an event to be $\mathbf{r},\mathbf{\theta}$ 
(for instance, these are values obtained in ring structure 
processing), the model of spatial-angular distribution of light is 
used to evaluate pattern image 
$\left \{ Q_j(E_0,\mathbf{r},\mathbf{\theta}),
\delta Q_j(E_0,\mathbf{r},\mathbf{\theta}) \right \}$
which is then used to calculate 
$H(\mathbf{r},\mathbf{\theta})$ (Eq.(19)) 
which compares it with 
the full light image of the event. Adjusting
$\mathbf{r}$ and $\mathbf{\theta}$
(see arrows in the figure) so that H approaches its minimum 
one obtains a better 
estimate of the true geometry parameters 
$\mathbf{r}_0,\mathbf{\theta}_0$.

%

\subsubsection{Application of the technique to simulated events}

By using the technique developed in the previous section, we analyze
simulated events to determine the error distributions for the vertex
position and for the particle direction.  Here, we examine the error
distributions for 300~MeV electrons and 500~MeV muons, which yield
roughly the same amount of Cherenkov light.

In Figure~15, we give the error distribution for the vertex position
for 300~MeV electrons for different Cherenkov threshold quantities.
``Ring only'' denotes that only the information from PMTs whose
Cherenkov photons contribute to the Cherenkov ring 
are used for the estimation of the error.  ``Full proc., thr=1ph'' 
denotes that information from ``ring only'' PMTs and also those
exceeding 1 Cherenkov photon are utilized for the estimation on the error. 
For ``full proc., thr=5'' and ``full proc., thr=10'', this latter threshold
is raised to 5 and 10 photons, respectively.

\begin{figure}
\rotatebox{90}{%
\resizebox{0.3\textwidth}{!}
{\includegraphics{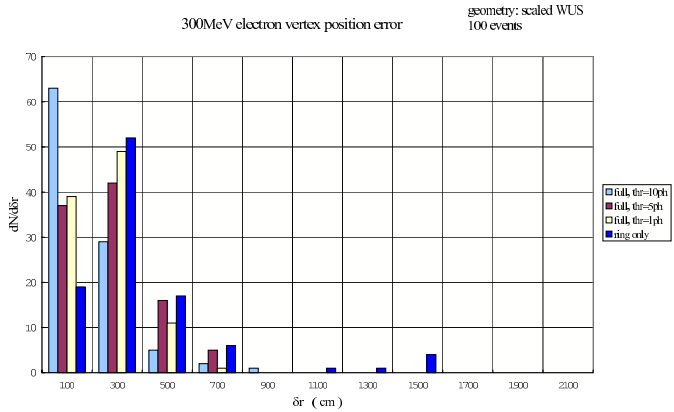}
}}
\caption{\label{fig:15}
Error distribution for the vertex position for 300 MeV 
electrons from the injected scaled WUS. 
The number of simulation is 100.
 }
\end{figure}

In Figure~16 the error distribution for the vertex position for 500~MeV 
muons are given. It is clear
that a wider error distribution is obtained for a ``ring only''
analysis.  A narrower error distribution results from the ``full proc,
thr=10 ph'' algorithm. This is the same as in the case of the
electron.  However, muons generally have wider error distributions
than electrons: the mean error for 500~MeV muons for the vertex
determination in the full analysis is 2.9~m while for 300 MeV electrons it amounts to 2 m.

\begin{figure}
\rotatebox{90}{%
\resizebox{0.3\textwidth}{!}{
  \includegraphics{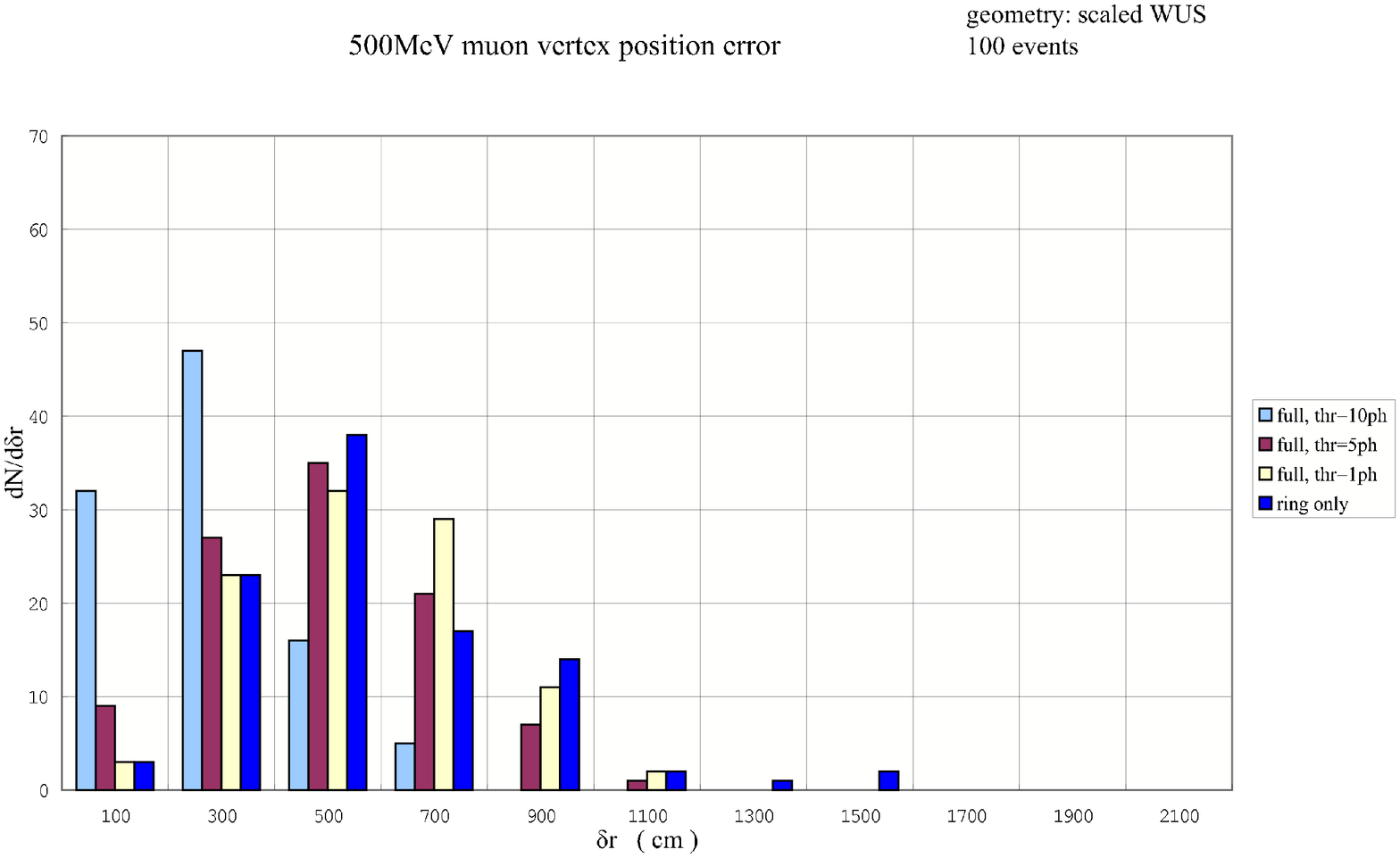}
}}
\caption{\label{fig:16} Error distribution for the vertex position for 500 MeV
muons from the injected scaled WUS. The number of simulation is100.}
\end{figure}

In Figure~17, the error distribution for the direction of the
300~MeV electron is given. As expected, ``ring only'' gives the largest
uncertainty distribution, while ``full proc., thr=10ph'' has the
narrowest error distribution, with a mean error of about 3.7$^\circ$.
In Figure~18,
 we give the corresponding distributions for muons. The
same trend is seen as for electrons, though the muons have a wider
uncertainty distribution.  The mean direction uncertainty in the best case is 4.9$^\circ$.

\begin{figure}
\rotatebox{90}{%
\resizebox{0.3\textwidth}{!}{
  \includegraphics{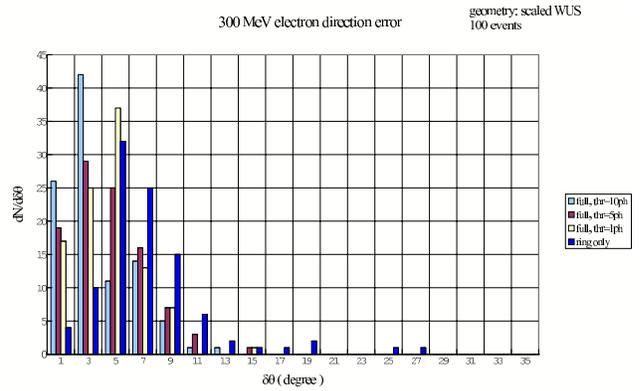}
}}
\caption{\label{fig:17} Error distribution for the direction for 300 MeV
 electrons from the injected scaled WUS. The number of simulation is 100.}
\end{figure}

\begin{figure}
\rotatebox{90}{%
\resizebox{0.3\textwidth}{!}{
  \includegraphics{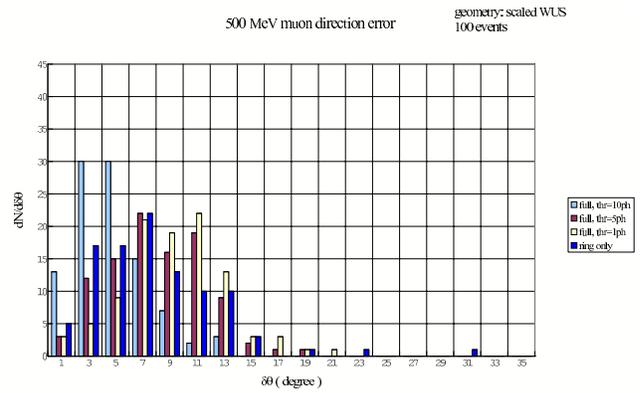}
}}
\caption{\label{fig:18} Error distribution for the direction for
 500 MeV muons from the injected scaled WUS.
 The number of simulation is 100.}
\end{figure}

Now, we compare 1~GeV electrons with 1~GeV muons, both of which yield
roughly the same quantity of Cherenkov light.  In
Figure~19, we give the error distribution for the vertex
position for 1~GeV electrons in the case of ``full proc,thr=10''. The
mean error is 1.9~m.  In Figure~20, the corresponding
quantities for the muon are plotted.  The average error for the vertex
position is 3.2~m.  Again, the error of the vertex point for muons is
larger than for electrons.

In Figure~21, the error distribution for the direction for
1~GeV electron is shown. The average direction error is 3.0$^\circ$
for ``full proc,thr=10''.  The corresponding quantities for the muon
are plotted in Figure~22, where the mean error is
5.3$^\circ$. Once again, the directional error for muons is larger
than that for electrons.

\begin{figure}
\rotatebox{90}{%
\resizebox{0.3\textwidth}{!}{
  \includegraphics{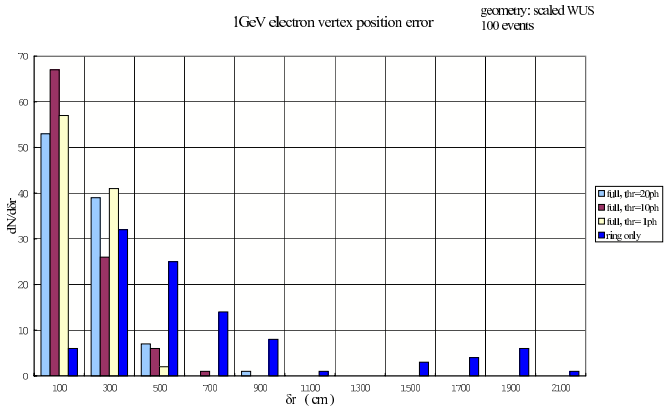}
}}
\vspace{-5mm}
\caption{\label{fig:19} Error distribution for the vertex position for 1 GeV
 electrons from the injected scaled WUS. The number of simulation is 100.}
\end{figure}

\begin{figure}
\rotatebox{90}{%
\resizebox{0.3\textwidth}{!}{
  \includegraphics{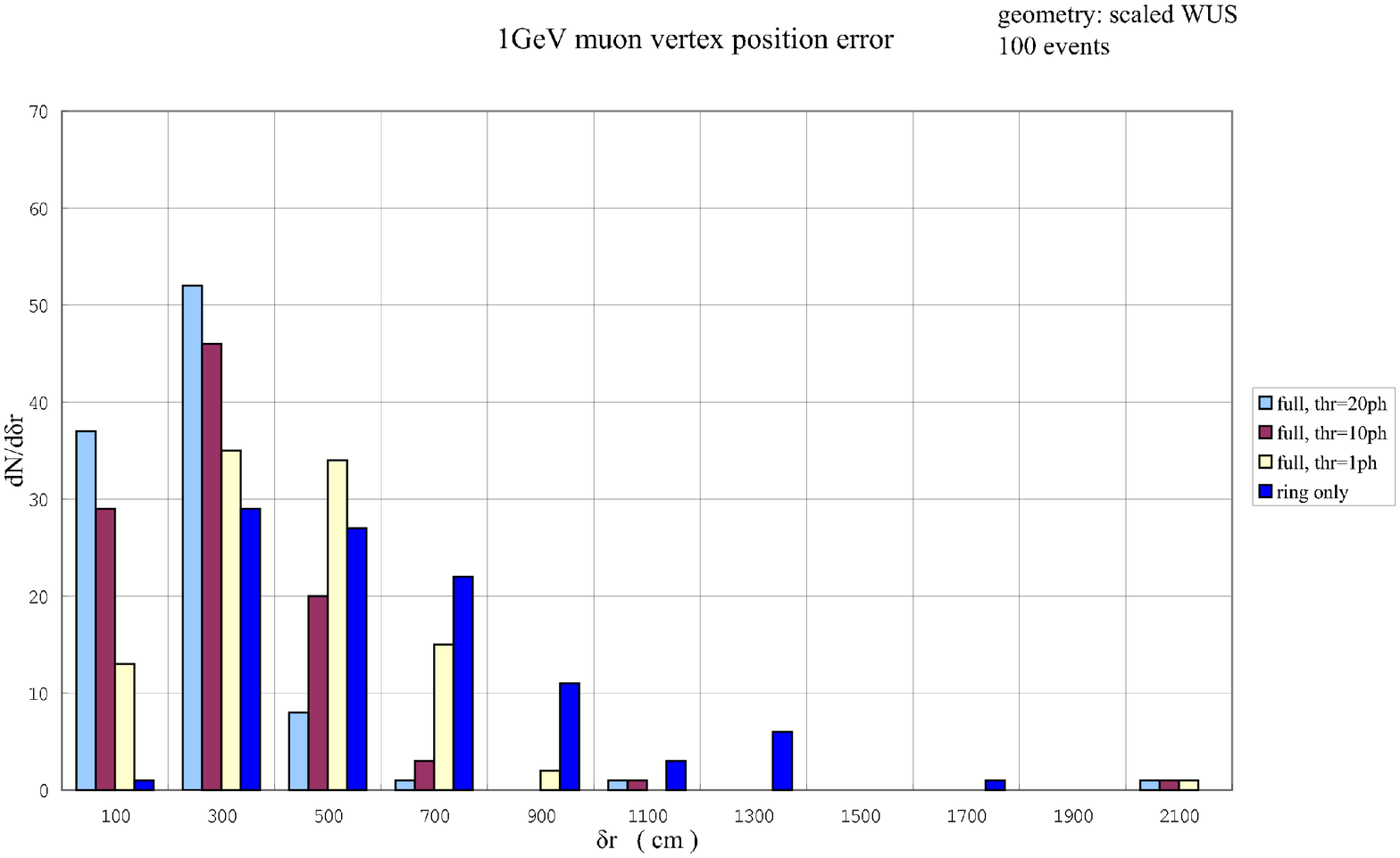}
}}
\vspace{-5mm}
\caption{\label{fig:20} Error distribution for the vertex position 1 GeV 
 muons from the injected scaled WUS. The number of simulation is 100.}
\end{figure}

\begin{figure}
\rotatebox{90}{%
\resizebox{0.3\textwidth}{!}{
  \includegraphics{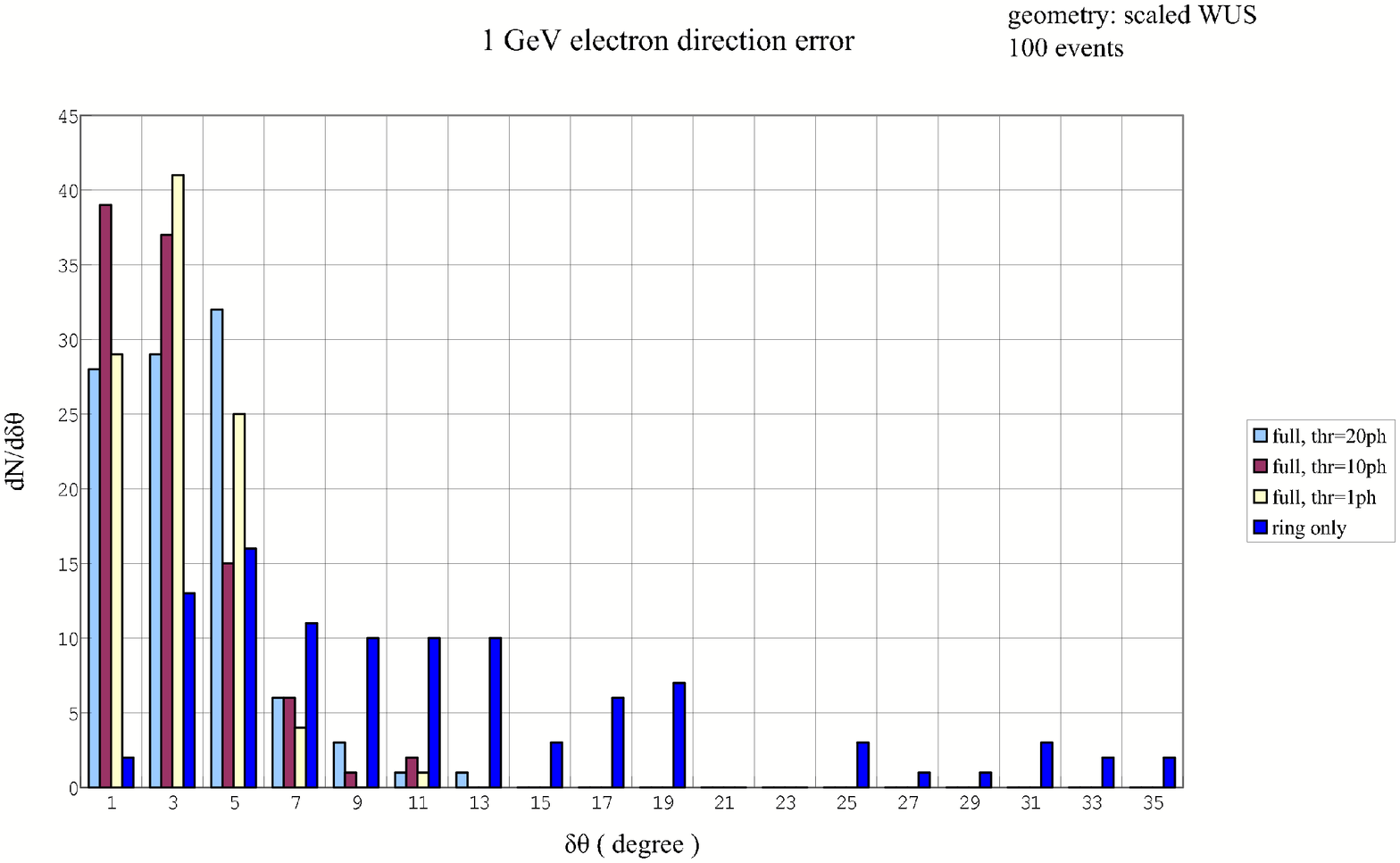}
}}
\caption{\label{fig:21} Error distribution for the direction for 1 GeV
 electrons from the injected scaled WUS. The number of simulation is 100.}

\rotatebox{90}{%
\resizebox{0.3\textwidth}{!}{
  \includegraphics{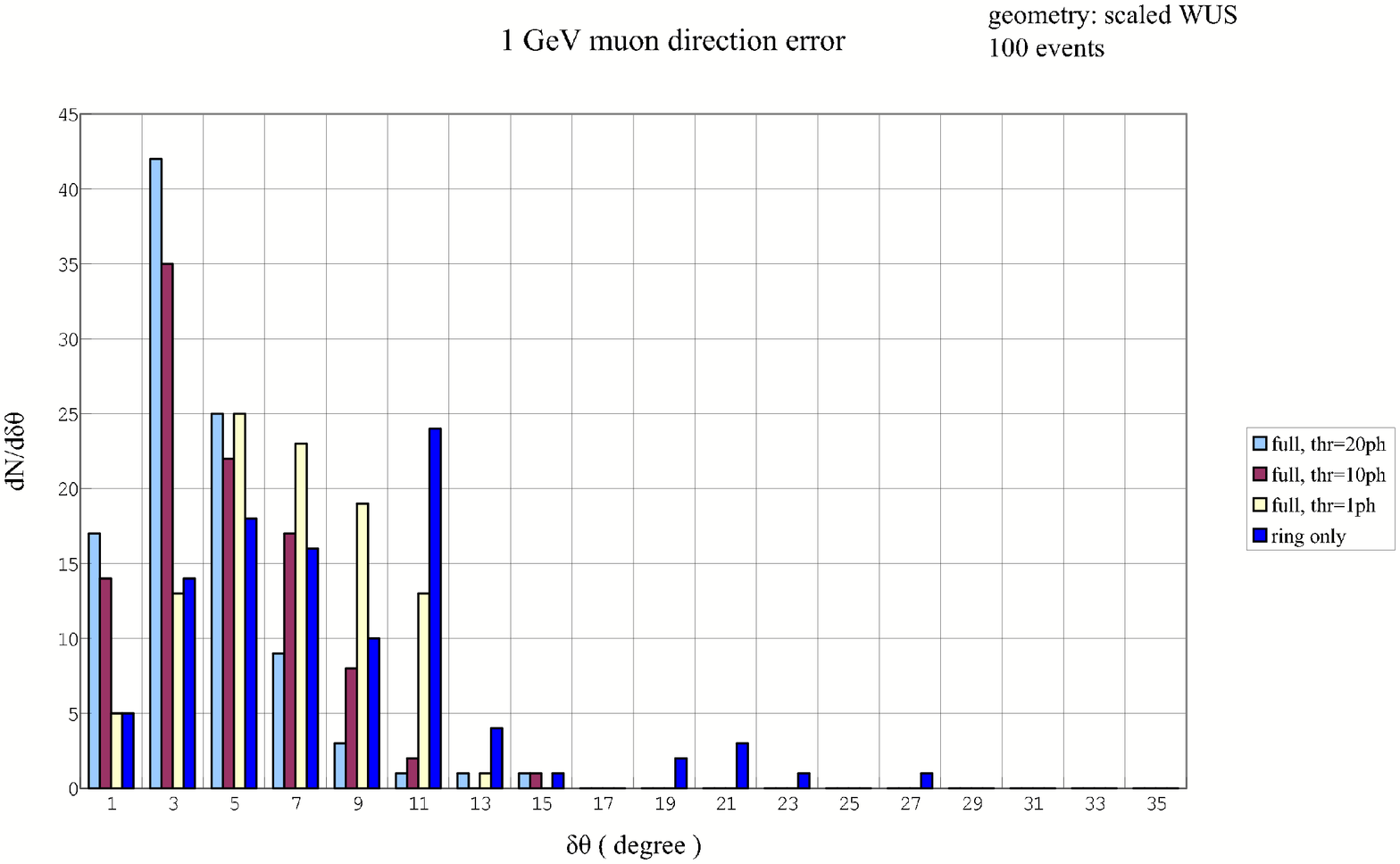}
}}
\caption{\label{fig:22} Error distribution for the direction 1 GeV 
 muons from the injected scaled WUS. The number of simulation is 100.}
\end{figure}

\begin{center}
\begin{table*}
\caption{\label{Table:3} 
Mean and standard deviation of the error in the vertex position and the
direction due to primary electrons and primary muons. These are given
 for different criteria for the Cherenkov threshold.  Ring proc. 
 denotes errors estimated using the Cherenkov 
 ring only. [1], [5], [10], [20] denote errors estimated by the 
 combination of [ring proc.] with a Cherenkov photon threshold 
 of 1, 5, 10 and 20photons, respectively.  Alm denotes the mean 
 direction error in degrees. Als denotes the standard deviation 
 for the corresponding mean values. Rm denotes the mean position
  error in metres. Rs denotes the standard deviation for the 
  corresponding mean values.
 }
\vspace{5mm}
\hspace*{4cm}
\begin{tabular}{|c|c|c|c|c|c|c|}
\hline
&&\multicolumn{5}{c|}{threshold}\\
\cline{3-7}
&&Ring proc.&[1]&[5]&[10]&[20]\\
\hline
         & Alm (deg.) &  7.2   &    4.4   &   4.6   &   3.7   &        \\
300 MeV  & Als (deg.) &  4.2   &    2.4   &   2.7   &   2.5   &        \\
electron &  Rm (m)   & 3.88   &    2.50  &   2.74  &   1.98  &        \\
         & Rs (m)    & 2.91   &    1.26  &   1.60  &   1.49  &        \\
\hline

         & Alm (deg.) &  7.8   &    9.3   &   8.0   &   4.9   &        \\
500 MeV  & Als (deg.) &  4.7   &    3.7   &   3.6   &   2.7   &        \\
muon     &  Rm (m)   &  5.71  &    5.69  &   4.92  &   2.94  &        \\
         &  Rs (m)   &  2.57  &    2.01  &   2.17  &   1.54  &        \\
\hline

         & Alm (deg.) & 11.7   &    3.1   &         &   3.0   &    3.7  \\
 1 GeV   & Als (deg.) & 14.4   &    3.5   &         &   3.7   &    4.4  \\
electron &   Rm (m)  &  6.49  &    1.99  &         &   1.86  &    2.17 \\
         &  Rs (m)   &  8.21  &    2.19  &         &   2.24  &    2.55 \\
\hline

         & Alm (deg.) & 8.8    &    7.3   &         &   5.3   &    4.9  \\
1 GeV    &Als (deg.)  &11.3    &   11.1   &         &  10.0   &   10.0  \\
muon     & Rm (m)    & 5.98   &    4.25  &         &   3.15  &    2.70 \\
         & Rs (m)    & 6.76   &    4.92  &         &   3.99  &    3.62 \\
\hline

         & Alm (deg.) &21.3    &          &         &   2.0   &    1.9   \\
5 GeV    &Als (deg.)  &11.0    &          &         &   5.1   &    4.4   \\
electron &  Rm (m)   &15.07   &          &         &   1.43  &    1.32  \\
         & Rs (m)    & 6.65   &          &         &   2.76  &    2.66  \\
\hline

         & Alm (deg.) & 8.5    &          &         &   4.3   &    3.8   \\
5 GeV    &Als (deg.)  & 6.7    &          &         &   4.0   &    3.8   \\
muon     & Rm (m)    & 4.68   &          &         &   2.89  &    2.45  \\
         & Rs (m)    & 2.52   &          &         &   2.55  &    2.45  \\

\hline
\end{tabular}
\end{table*}
\end{center}

In Table~3, we summarize the error distributions for the vertex points
and the direction for both electrons and muons.  Both mean errors and
root mean square errors are given.  Errors are also given for the
different criteria, namely, different Cherenkov light threshold. 
 From Figures~15 to 18 and Figures~19 to 22 and Table~3, 
it should be noticed the followings:

\begin{enumerate}

\item Of the different criteria considered, the ``ring only''
procedure results in the largest error.  The reasons are as follows:
The concept of the ring structure is essentially fuzzy, both in our
procedure and the SK procedure, and information from ring structure is
only part of the total information available for the pattern
recognition.  It is, therefore, natural that the vertex position and
directional errors are largest in the ``ring only'' analysis.  The
standard SK analysis uses ring structure only, and their errors are
amplified by that fact that the analysis ignores fluctuation effects.

\item Muons events have larger uncertainties than electron events.
For both electrons and muons, the sources for the Cherenkov light are
not point-like and have some extent in both cases. Significant errors
come from the point-like approximation for electron events.

\item The optimal Cherenkov threshold for the third step of geometry 
reconstruction procedure depends on the primary energy of the particle 
concerned. For energies less than 1 GeV, third step with 10 photon 
threshold gives the best results for both  muons and electrons 
among the alternatives considered. For 5 GeV electrons and muons,
 20 photon threshold seems to be optimal for the third step.
 
\item The fact that the uncertainties for the determination of the
vertex point and direction are rather large comes from the effect of
fluctuations, namely the nature of the stochastic process concerned
(an electron cascade shower or sequence of muon interactions with the
medium).  The utility of model developed in this paper, the moving
point approximation model, is guaranteed, because it gives mean values
and relative fluctuations precisely and takes all necessary
geometrical considerations into account correctly.  Even if additional
errors exist, they should be negligible compared to the uncertainty
caused by fluctuations.  The rather large errors for the vertex point
and the direction obtained by our model could not be reduced substantially,
reflecting the essential nature of the physical
processes concerned.

\item The SK analyses, according to all published accounts, completely
neglect fluctuations and also use point-like approximations for the
electron cascade.  Moreover, they neglect the scattering
effects on muon track geometry. As we showed earlier, their simple
approximations distort the mean values in certain parameter
domains. The most probable reason for their low error estimates is the
fact that they completely neglect fluctuations in the event
development.  Our results contradict clearly the fine positional
resolution of 23 to 56\,cm claimed by for the SK analysis
(Kibayashi, p.73\cite{r7Kibayashi}).

\item It should be noticed that errors derived by us are lower limits.
As already mentioned, we do not consider the production of
photoelectrons in PMTs, and only consider direct Cherenkov photons in our
discrimination procedures neglecting the diffusion of
Cherenkov photons. If we include these factors in our procedure, then,
the actual errors for the vertex position and the direction should be
larger than that given here.

\item Here, we make remarks to the results obtained by Mitsui 
et al(\cite{r8Mitsui}. They have calculated photoelectrons 
from the Cherenkov light due to 
both muon-like events and electron like events in the SK experiment by full 
Monte Carlo Method and have showed the clear deviation of the photoelectrons
 concerned from the Poisson distribution which SK assume(See,Eq.(11) in the 
 present paper). Further, they have calculated PID parameter values following
the SK PID methods which consists of the pattern and opening angles of
the Cherenkov light. As the resluts of it, they have concluded that 
the total misidentification for muon and electron events is larger 
than or equal to 20 percent in sub-Gev and also at several percent 
in multi-GeV region.
  They attribute the deviation from the Poisson distribtuion to the 
neglect of the fluctuation from the stochastic proceeses inherent 
in electron-like eventsand muon-like events. We agree to their 
interpretation. However, we are forced to criticize the adoption 
of the same PID ESTIMATOR as the SK's. We have 
already pointed out the inadequency of SK PID ESTIMATOR ( See Galkin
\cite{r1}, page 6).

\end{enumerate}
%

\section{Summary}

\noindent(1) The SK TDC procedure 

The TDC procedure assumes that the Cherenkov light originates from a
point, and thus does not determine the vertex position accurately,
because the sources for the Cherenkov light have a non-negligible
extent.  In order to utilize the TDC meaningfully, we should take into
account the extent of the source for the Cherenkov light in space and
time. Further, ideally we should utilize not only arrival time of the
Cherenkov light but also shape of the pulse in the PMTs.\\

\noindent(2) The SK ADC procedure: the estimator for particle identification

It is no exaggeration to state that the credibility of the
discrimination of muon events from electron events in the SK analysis
is entirely dependent on the validity of the estimator adopted by the
SK.

The estimator for particle identification was introduced into the
analysis of the neutrino events in the Kamiokande detector 
(Takita\cite{r2Takita}).  The main purpose of the KEK 1~kilo-ton experiment
was to test, using particles of known type and momentum, the validity
of the Monte Carlo simulation used in the Kamiokande detector (Kasuga, 
p.22\cite{r3Kasuga}).  The results of this test established the
validity of the estimator and thus it was applied to the standard SK
analysis (Kasuga et al.\cite{r4Kasuga}, Sakai\cite{r5Sakai}, Kasuga
\cite{r3Kasuga}).  As a result of using this estimator, it has been asserted
that evidence for neutrino oscillations between muons and tauons has
been obtained.

As we have pointed out, the SK estimator for particle identification
consists of three parts. The first is the probability function, in
which the spatial part is separated from the angular part.  The second
is the probability function for photoelectron production.  The third
is the introduction of the point-like approximation into the Cherenkov
light for electrons.  As we have made clear, the first
two lack a firm theoretical foundation. The third is an
oversimplification which leads generally to a wrong estimation for the
quantity of Cherenkov light.  Consequently, the SK estimator for
particle identification can not be reliable for an accurate
discrimination between electrons and muons.  SK results depending on
the discrimination of electron-like events from muon-like events (see
Kasuga et al.\cite{r4Kasuga}) must be re-evaluated in this light.

\begin{figure}
\begin{center}
\resizebox{0.5\textwidth}{!}{%
  \includegraphics{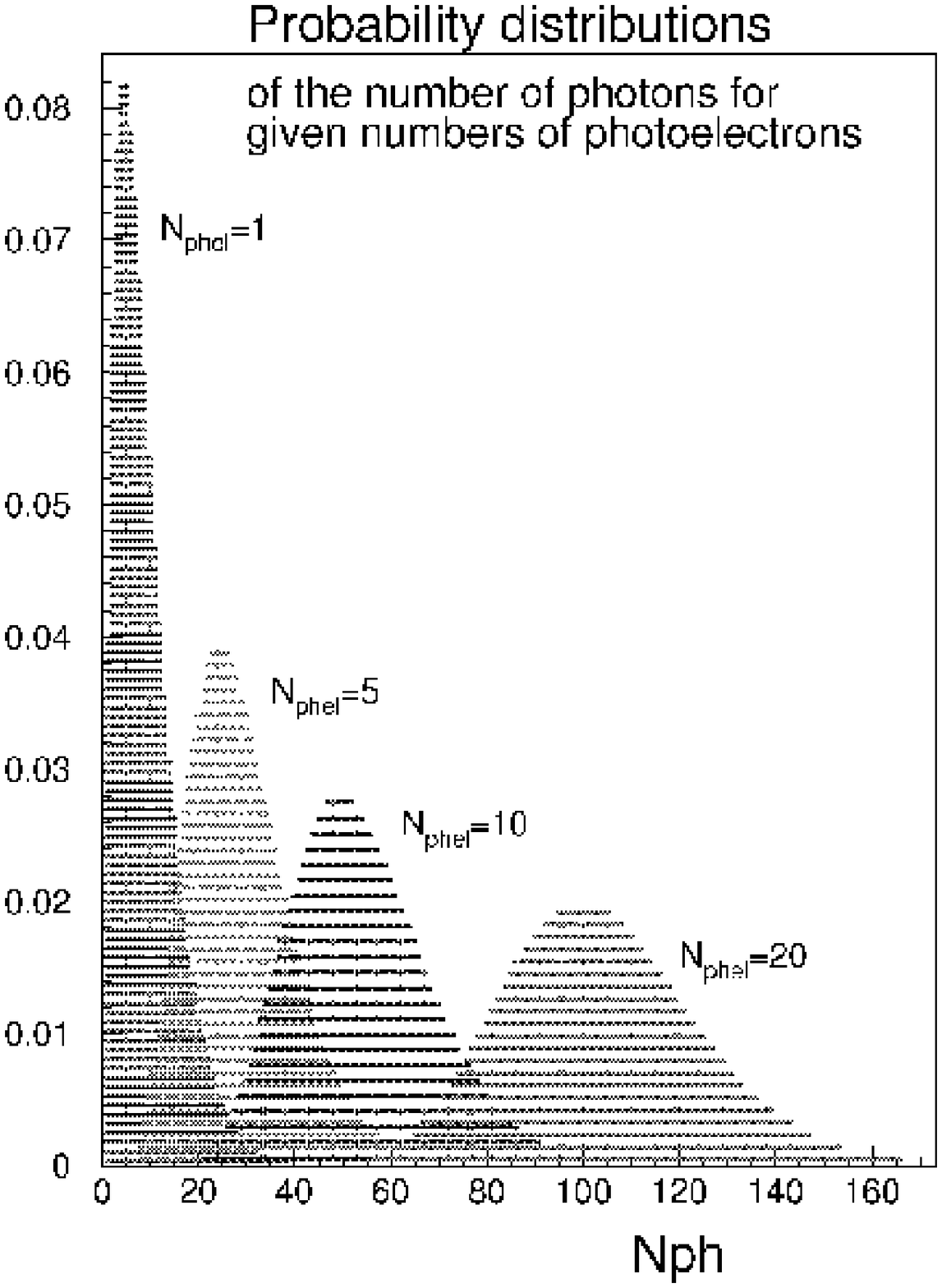}
}
\caption{\label{fig:23}
 Probability distributions of the number of the Cherenkov photons 
for given numbers of photoelectron. The number of simulation is 100.}
\end{center}
\end{figure}

\noindent(3) Errors for the vertex point and the direction

Only the correct estimator for particle identification can estimate
the error for the vertex point and direction quantitatively. Our
estimation (Figures~15 to 18 and Figures~19 to 22 and Table~3) 
shows non-negligible and
inevitable errors for the vertex position and direction.  It should
be, particularly, noticed that the fluctuations in error in both
vertex position and direction are too big.

 Kibayashi (p.73\cite{r7Kibayashi}) concludes that the uncertainty for
vertex point is from 23~cm to 56~cm and the uncertainty for the
direction from 0.9$^\circ$ to 3.0$^\circ$ using both the estimator for
particle identification and the TDC adopted by the SK.  As we have
demonstrated, these appear to severely underestimate the error
distributions, which is too far from reality.
\\
\noindent (4) In the present paper, we do not take into consideration photoelectrons
produced by the Cherenkov light in the PMT, and we neglect scattering
of the Cherenkov light. Therefore, our results on
discrimination of electron events from muon events, only yield 
lower
limits to the realistically achievable experimental 
erroes. 
  If we
consider these factors, the situation becomes much more complicated.
In Figure~23,
 we give the Cherenkov light distributions for given
numbers of photoelectron.  It is easily understood from the figure
that the patterns of electron events and muon events are much more
vague, because there is non-negligible distribution of the
photoelectron for a given number of Cherenkov photons.  Further, we
can say that it is quite sufficient for us to limit our analysis to
the level of direct Cherenkov light, neglecting scattered Cherenkov
light, to clarify the essential points in the discrimination procedure
between electrons and muons.
\\
\noindent(5) As a result of the points mentioned in (4), our results give lower
limits for the primary parameter uncertainties in  
the problems concerned. We show a clear separation between
electron and muon by using a procedure based on detailed Cherenkov
light angular distribution approximations.  Even though we show a
clear separation between electron and muon for the same quantity of
the Cherenkov light, this remains at the theoretical level and does
not provide the optimal experimental means of separating electrons
from muons. This problem is beyond the scope of this paper and is
better undertaken by those with a detailed understanding of the
specifics of the detector.

A part of the preceding paper\cite {r1} and present paper are found
in \cite {r9Anokhina}.

\section{Acknowledgement}


One of the authors (V.G.) should like to thank Prof.\ M.\ Higuchi,
Tohoku Gakuin University. Without his invitation, V.G.\ could not join
in this work.  Authors would like to be very grateful for the
remarkable improvement of the manuscript to Dr. Philip Edwards.

%

\begin{thebibliography}{}
%
%
\bibitem{r1} Galkin,V.I.,Anokhina,A.M.,Konishi,E.~and~Misaki,A.,
 Eur.Phys.J.C(submitted), arXiv:hep-ex/0412059
\bibitem{r2Takita} Takita,M., PhD thesis, University of Tokyo (1988)
\bibitem{r3Kasuga} Kasuga,S., ICRR Report~338-95-4 (1995)
\bibitem{r4Kasuga} Kasuga,S. {\it et al.}, Phys.Lett.{\bf B374}, (1996) 238 
\bibitem{r5Sakai} Sakai,A., PhD thesis, University of Tokyo (1997)
\bibitem{r6Kasuga2} Kasuga,S., PhD thesis, University of Tokyo (1998)
\bibitem{r7Kibayashi} Kibayashi,A., PhD thesis, University of Hawaii (2002)

\bibitem{r8Mitsui} Mitsui,K.,Kitamura,T.,Wada,T.~and~Okei,K.,
 J.Phys.G:Nucl.Part Phys.{\bf 29 }(2003)2281
\bibitem{r9Anokhina} Anokhina,A.M. and Galkin,V.I.,Physics of Atomic Nuclei {\bf 69} No.1 (2006) 16 
\end{thebibliography}
%

\end{document}